\documentclass[
    aps,
    pra,
    twocolumn,
    superscriptaddress,
    longbibliography,
    nofootinbib
]{revtex4-2}

\usepackage[english]{babel}
\usepackage{amsmath,amsfonts,amssymb}
\usepackage{ulem}
\usepackage{physics}
\usepackage{bm}
\usepackage{amsmath}
\usepackage{amsthm}
\usepackage{xcolor}
\usepackage[most]{tcolorbox}

\usepackage{graphicx}
\usepackage{tikz}
\usepackage{placeins}
\usepackage{subcaption}

\usepackage{algorithmicx}
\usepackage{algcompatible}
\usepackage{float}

\newcommand{\Sbar}{\bar{\boldsymbol{S}}}

\newtheorem{proposition}{Proposition}
\newtheorem{corollary}[proposition]{Corollary}
\newtheorem{lemma}[proposition]{Lemma}

\newcommand{\uv}{Departament de Física Teòrica and IFIC, Universitat de València-CSIC, 46100 Burjassot (València), Spain}

\newcommand{\upv}{Departament de Matemàtica Aplicada, Universitat Politècnica de València, València, Spain}

\newcommand{\iumpa}{Instituto Universitario de Investigación de Matemática Pura y Aplicada, Universitat Politècnica de València, València, Spain}

\newcommand{\ceu}{Departamento de Matemática, Física y Ciencias Tecnológicas, Universidad CEU Cardenal Herrera, València, Spain}

\usepackage[title]{appendix}
\usepackage[colorlinks]{hyperref}
\usepackage{lipsum}

\begin{document}

\title{Sparse-QITE: Discovering Compact Generators for Quantum Imaginary-Time Evolution via Regularization Methods}

\author{Rafael Gómez-Lurbe}
\thanks{These authors contributed equally to this work and share first authorship.}
\affiliation{\uv}
\email{Corresponding author: rafael.gomez-lurbe@uv.es}

\author{Luca Ion}
\thanks{These authors contributed equally to this work and share first authorship.}
\affiliation{\uv}

\author{Alberto Acevedo}
\affiliation{\ceu}
\affiliation{\iumpa}

\author{Andreu Anglés-Castillo}
\affiliation{\uv}
\affiliation{\upv}

\begin{abstract}
Quantum imaginary-time evolution (QITE) provides a systematic route to ground-state preparation, but its cost grows rapidly with the locality \(D\) of the unitary approximation, for which the local Pauli basis scales as \(4^D\). We introduce Sparse-QITE, a regularized formulation that promotes compact generators, reducing the number of quantum operations and yielding circuits with fewer gates and lower depth. We derive rigorous bounds on the sparsification error and benchmark the method in one-dimensional transverse-field Ising and antiferromagnetic Heisenberg models. Sparse-QITE exhibits a controllable sparsity--accuracy trade-off while retaining strong compression even in the high-accuracy regime. In scalability experiments with respect to \(D\), the active support grows much more slowly than in standard QITE and, in several regimes, approximately saturates over the domain sizes studied. After circuit construction and transpilation, some of the largest-domain cases require only about \(1\%\) of the standard-QITE gate count while maintaining high accuracy. We further introduce adaptive and active-discovery variants: the former selects the regularization strength dynamically during the evolution, while the latter reduces the number of Pauli expectation values that need to be evaluated and, consequently, the measurement requirements. These results establish sparsity-promoting regularization as a route toward more resource-efficient QITE. More broadly, these results suggest that sparsity-promoting regularization may also benefit variational quantum algorithms.
\end{abstract}

\maketitle

\section{Introduction}

Quantum simulation is a central motivation for quantum
computing~\cite{Feynman1982,Cirac2012,mindgaps}. Ground-state
preparation~\cite{Bauer2020} is particularly relevant to quantum
chemistry~\cite{chem1,chem2}, materials
science~\cite{Auerbach1994,Sachdev2011}, and optimization~\cite{IsingNP}.
Although powerful classical methods exist for many
regimes~\cite{RevModPhys.93.045003,Suzuki1976}, their limitations motivate
quantum approaches that represent many-body states directly on quantum
hardware.

Variational approaches, including VQE and
QAOA~\cite{Peruzzo2014,QAOA}, use parametrized circuits whose performance
depends on the ansatz and optimization strategy, motivating problem-tailored architectures and geometry-aware
updates~\cite{Romero2018,geodesicansatz,Stokes2020quantumnatural}. Quantum phase estimation offers another route to ground-state preparation,
but generally requires deeper circuits.

Imaginary-time evolution (ITE) provides another route to ground-state
preparation by progressively suppressing excited-state components, provided
the initial state has nonzero ground-state overlap. Classical approaches
based on tensor networks~\cite{Schollw_ck_2011,Ba_uls_2023}, quantum Monte
Carlo~\cite{RevModPhys.73.33,Assaad2008}, and Pauli
propagation~\cite{G_mez_Lurbe_2026} approximate this dynamics in different
tractable regimes. Implementing ITE on quantum hardware is challenging
because it is nonunitary. Quantum imaginary-time evolution
(QITE)~\cite{Motta2019} addresses this difficulty by approximating each short
imaginary-time step with a unitary transformation on a local domain, without
requiring a global variational ansatz.

In standard QITE, each local unitary is determined by expanding its generator
in a Pauli basis and solving a linear system for the coefficients. For a
domain of \(D\) qubits, the full basis contains \(4^D\) strings, leading to rapidly growing measurement and classical-processing requirements.
Implementing generators with many active terms also increases gate counts
and circuit depth. Since the domain needed for an accurate approximation
depends on correlations in the evolving state~\cite{Motta2019}, restricting
\(D\) can limit accuracy. Reducing the resources associated with larger
domains is therefore an important challenge.

Existing approaches address the cost of imaginary-time evolution through,
for example, double-bracket constructions~\cite{DBQITE} or the geometric
structure of the trajectory~\cite{ACQ}. Here, we investigate a complementary
question: must the full local Pauli space be populated to reproduce QITE
dynamics accurately? We introduce \textit{Sparse-QITE}, which uses
sparsity-promoting regularization~\cite{1614066} to select compact generators
from the available operator pool. An \(\ell_1\) penalty promotes vanishing
coefficients, while an additional \(\ell_2\) term stabilizes their
determination, yielding an elastic-net formulation. The regularization strength controls the balance between generator sparsity and approximation
accuracy.

We derive rigorous bounds quantifying the trajectory error between Sparse-QITE and ordinary QITE. Numerical benchmarks in the transverse-field
Ising model, in both ordered and disordered phases, and the antiferromagnetic
Heisenberg model demonstrate substantial generator compression, including
regimes with ground-state accuracy comparable to standard QITE.
Importantly, the active support grows much more slowly than the full Pauli
pool over the domain sizes studied and approximately saturates in several
regimes. Although these results do not establish an asymptotic scaling law,
they indicate that increasing the domain need not entail a corresponding
increase in the number of implemented generator terms. This is particularly
relevant when the domain required for convergence is not known in advance. The compression persists after circuit construction and transpilation,
producing substantial reductions in gate counts and circuit depth.

We further introduce two extensions. \textit{Adaptive Sparse-QITE} selects
the regularization strength at each update according to a local
approximation tolerance, providing additional compression at matched
trajectory accuracy. Basic Sparse-QITE still requires the expectation
values associated with the full candidate pool;
\textit{Active-Discovery Sparse-QITE} addresses this remaining cost by
restricting subsequent updates to support identified during an initial
discovery stage. This reduces the number of required Pauli expectation
values, with an accuracy cost that depends on the regime. Together, these
methods exploit generator sparsity to reduce circuit resources and, through
support reuse, measurement requirements. 

\begin{figure*}[!t]
    \centering
    \includegraphics[width=0.95\textwidth]{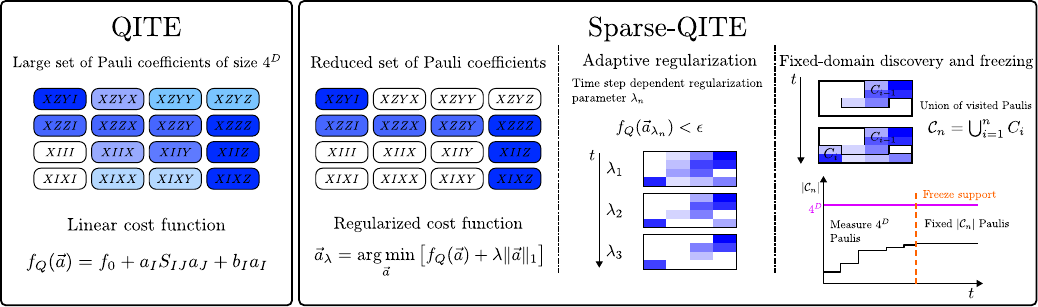}
    \caption{
Overview of the Sparse-QITE framework. Sparse regularization selects compact
Pauli generators; Adaptive Sparse-QITE adjusts the regularization strength
according to a local approximation tolerance; Active-Discovery Sparse-QITE
reuses discovered support to reduce subsequent measurement requirements.
    }
    \label{fig:SpQite_Schema}
\end{figure*}

Although we use standard QITE as a baseline, the same \(\ell_1\)-regularization strategy can be incorporated into QITE-derived methods that minimize an objective, and more broadly into variational formulations where sparse parameters translate into reduced implementation costs. Figure~\ref{fig:SpQite_Schema}
summarizes the Sparse-QITE framework.


\section{Background}\label{sec:background}

In this section, we present the main ingredients underlying Sparse-QITE. We first introduce imaginary-time evolution and its use for ground-state preparation. We then describe the standard QITE construction, focusing on the local unitary approximation, the associated linear system, and the scaling of its candidate Pauli set with the domain size. Finally, we briefly review sparse regularization methods from compressed sensing, which provide the optimization framework used to construct compact QITE generators.

\subsection{QITE}

The core idea of QITE methods is to approximate a Trotterized imaginary time evolution (which is non-unitary) by unitary operations
\begin{equation}\label{eq:QITE_cond}
    \frac{e^{-\Delta\tau\boldsymbol{\hat{h}}_k}\ket{\psi_n}}{\norm{e^{-\Delta\tau\boldsymbol{\hat{h}}_k}\ket{\psi_n}}} \sim e^{-i\Delta\tau \boldsymbol{\hat{A}}_{n,k}}\ket{\psi_n}~,
\end{equation}
where $\boldsymbol{\hat{H}} = \sum_k^{N_H} \boldsymbol{\hat h}_k$ is the problem Hamiltonian whose ground state we are interested in, $\ket{\psi_n}$ is the quantum state at a discrete time step $n$,  and $\boldsymbol{\hat{A}}_{n,k}$ is the unitary generator that approximates ITE generated by the Hamiltonian piece $\boldsymbol{h}_k$. The construction of the generator is performed by expressing it in a Hermitian basis, for instance the basis of Pauli strings 
\begin{equation}
	\boldsymbol{\hat{A}_n} = \sum_I a_{I}^{(n,k)} \boldsymbol{\hat{\sigma}}_I~,
\end{equation}
where the $\sigma_I$ represent the collection of Pauli strings included in the local generator, and finding the real coefficients $a_{I}^{(n,k)}$ that minimize the state distance between the states in Eq.~\eqref{eq:QITE_cond}. We refer to the set of Pauli strings included in this expansion as the candidate Pauli set. This condition sets the objective function to be minimized
\begin{equation}\label{eq:QITE_objective}
	f(a_I) = f_0 + \sum_I b_I a_I + \sum_{I,J} a_I S_{IJ} a_J~,
\end{equation}
where $f_0$ is a constant factor depending on the ITE state, and the remaining terms
\begin{align}
S_{IJ}& = \expval{\boldsymbol{\hat{\sigma}}_I\boldsymbol{\hat{\sigma}}_J}{\psi_n}  ~,\\
b_I& = \frac{i}{c\Delta \tau}\expval{[e^{-\boldsymbol{\hat{h}}_k\Delta \tau},\boldsymbol{\hat{\sigma}}_I]}{\psi_n} ~,\\
	c &= \sqrt{\expval{e^{-2\boldsymbol{\hat{h}}_k\Delta\tau}}{\psi_n}}~,
\end{align}
which involve expectation values on the state $\ket{\psi_n}$ and we dropped the explicit time step $n$ and Hamiltonian piece $k$ dependence of $a_I$, $S_{IJ}$ and $b_I$ to alleviate notation, and will be dropped when not necessarily required.

Because the coefficients $a_I$ are real, as required for $\boldsymbol{\hat{A}}_{n,k}$ to be Hermitian and hence for the update in Eq.~\eqref{eq:QITE_cond} to be unitary, only the symmetric part of $\boldsymbol{S}$ contributes to the quadratic term of Eq.~\eqref{eq:QITE_objective}. Indeed, decomposing \(\boldsymbol{S}=\tfrac{1}{2}\bigl(\boldsymbol{S}+\boldsymbol{S}^{T}\bigr)+\tfrac{1}{2}\bigl(\boldsymbol{S}-\boldsymbol{S}^{T}\bigr)\), the antisymmetric part drops out, since \(\vec{a}^{T}\bigl(\boldsymbol{S}-\boldsymbol{S}^{T}\bigr)\vec{a}=\vec{a}^{T}\boldsymbol{S}\,\vec{a}-\bigl(\vec{a}^{T}\boldsymbol{S}\,\vec{a}\bigr)^{T}=0\) for any real \(\vec{a}\). The stationarity condition \(\nabla f=0\) therefore gives the QITE linear system~\cite{Motta2019}
\begin{equation}
\label{eq:QITE_linear_system}
    \bigl(\boldsymbol{S}+\boldsymbol{S}^{T}\bigr)\vec{a}=-\vec{b}.
\end{equation}
Since $\boldsymbol{S}$ is Hermitian, $\boldsymbol{S}^{T}=\boldsymbol{S}^{*}$. We therefore define the symmetrized Gram matrix
\begin{equation}
\label{eq:real_gram}
\begin{split}
    \Sbar&:=\tfrac{1}{2}\bigl(\boldsymbol{S}+\boldsymbol{S}^{T}\bigr)=\operatorname{Re}\boldsymbol{S},\\
    \Sbar_{IJ}&=\tfrac{1}{2}\expval{\{\boldsymbol{\hat{\sigma}}_I,\boldsymbol{\hat{\sigma}}_J\}}{\psi_n},
\end{split}
\end{equation}
which is real, symmetric and positive semidefinite. For real $\vec{a}$ one has $\vec{a}^{T}\boldsymbol{S}\,\vec{a}=\vec{a}^{T}\Sbar\,\vec{a}$, and Eq.~\eqref{eq:QITE_linear_system} reads $2\Sbar\vec{a}=-\vec{b}$. It is therefore the spectrum of $\Sbar$, rather than that of $\boldsymbol{S}$, that controls the conditioning of the coefficient determination.

In practice, the Hamiltonian is Trotterized into local pieces acting on \(T\) neighbouring spins, while the Pauli strings entering the generator are restricted to a local domain of \(D\geq T\) spins. This common practice introduces a Trotter error of the same order as the Trotterized evolution while reducing the resource requirements of the algorithm. Nevertheless, the candidate Pauli set on a domain of \(D\) spins can contain up to \(4^D\) Pauli strings, giving rise to two main bottlenecks: the measurement cost associated with estimating the quantities entering the QITE linear system, and the circuit complexity associated with implementing the unitary generated by \(\boldsymbol{\hat{A}}_{n,k}\). Since the synthesis cost of this unitary is directly related to the number of non-zero coefficients \(a_I\), this motivates modifying the objective function in Eq.~\eqref{eq:QITE_objective} to simultaneously approximate ITE and promote sparsity in the generator. In this work, we focus primarily on reducing this circuit complexity, while also introducing a variant that exploits the discovered sparse support to reduce the measurement requirements.

\subsection{Sparse regularization}
\label{sec:sparse_regularization}

Regularization methods are widely used in optimization, machine learning and compressed sensing to control the complexity and stability of a solution~\cite{hastie2009elements,1614066}. In general, regularization modifies an optimization problem by adding a penalty that constrains the magnitude or structure of the parameters. For a parameter vector \(\mathbf{a}\), a regularized problem can be written as
\begin{equation}
    \min_{\vec{a}}
    \left[
        f(\vec{a})
        +
        \lambda R(\vec{a})
    \right],
\end{equation}
where \(f(\vec{a})\) is the original objective function, \(R(\vec{a})\) is the regularization penalty, and \(\lambda \geq 0\) controls the strength of the regularization.

To directly promote sparsity, one would ideally penalize the number of nonzero coefficients through the \(\ell_0\) pseudonorm,
\begin{equation}
    \|\vec{a}\|_0
    =
    \#\{i : a_I \neq 0\}.
\end{equation}
Minimizing an objective function containing this penalty, however, generally leads to a combinatorial and NP-hard optimization problem~\cite{natarajan1995sparse}.

A computationally convenient alternative is \(\ell_2\)-regularization, also known as ridge or Tikhonov regularization, for which
\begin{equation}
    R(\vec{a})
    =
    \|\vec{a}\|_2^2
    =
    \sum_I |a_I|^2.
\end{equation}
This penalty yields a convex problem that can often be solved efficiently and can improve the conditioning and stability of the solution~\cite{tikhonov1977solutions}. Nevertheless, \(\ell_2\)-regularization typically shrinks the coefficients continuously without setting them exactly to zero, and therefore generally produces dense solutions. This is undesirable in the present setting, where the objective is not only to reduce the magnitudes of the Pauli coefficients but also to eliminate Pauli strings from the QITE generator.

A suitable compromise between sparsity and computational tractability is provided by \(\ell_1\)-regularization,
\begin{equation}
    R(\vec{a})
    =
    \|\vec{a}\|_1
    =
    \sum_I |a_I|.
\end{equation}
The \(\ell_1\) penalty preserves convexity while promoting solutions in which many coefficients vanish exactly~\cite{tibshirani1996regression,candes2008introduction}.

This sparsity-inducing behavior follows from the nonsmoothness of the absolute value at the origin. Consider the convex optimization problem
\begin{equation}
    \min_{\vec{a}}
    \left[
        f(\vec{a})
        +
        \lambda \|\vec{a}\|_1
    \right],
\end{equation}
where \(f(\vec{a})\) is differentiable and convex. The subdifferential of the absolute value is
\begin{equation}
\partial |a_I| =
\begin{cases}
\{1\}, & a_I > 0,\\
[-1,1], & a_I = 0,\\
\{-1\}, & a_I < 0.
\end{cases}
\end{equation}
The corresponding optimality conditions are
\begin{equation}
\begin{cases}
\dfrac{\partial f}{\partial a_i}(\vec{a}^*)
=
-\lambda\,\operatorname{sign}(a_I^*),
& a_I^* \neq 0,\\[3mm]
\left|
\dfrac{\partial f}{\partial a_I}(\vec{a}^*)
\right|
\leq \lambda,
& a_I^* = 0.
\end{cases}
\end{equation}
Thus, a zero coefficient is compatible with optimality whenever the corresponding gradient component lies within the finite interval \([-\lambda,\lambda]\). In contrast, for an \(\ell_2\) penalty the optimality condition is
\begin{equation}
    \frac{\partial f}{\partial a_I}(\vec{a}^*)
    +
    2\lambda a_I^*
    =
    0.
\end{equation}
At \(a_I^*=0\), this requires the more restrictive condition
\begin{equation}
    \frac{\partial f}{\partial a_I}(\vec{a}^*)
    =
    0.
\end{equation}
The finite interval allowed by the \(\ell_1\) optimality condition is therefore the mechanism through which coefficients can be driven exactly to zero.

When applied to a least-squares objective, \(\ell_1\)-regularization is commonly known as the least absolute shrinkage and selection operator (LASSO)~\cite{tibshirani1996regression}. In this work, we apply this approach to the QITE optimization problem to obtain generators containing fewer active Pauli strings. The resulting formulation, which we call \textit{Sparse-QITE}, is introduced in the following section.

\section{Sparse-QITE}
\label{sec:sparse_qite}

As reviewed in Sec.~\ref{sec:background}, standard QITE determines, at each imaginary-time step, a Hermitian generator whose associated unitary approximates the corresponding nonunitary evolution. For a given Hamiltonian piece $h_k$ and a domain of \(D\) qubits, let \(\mathcal P_{n,k}\) denote the candidate Pauli set used to expand the generator. In the standard QITE construction, \(\mathcal P_{n,k}\) may be chosen as the full set of Pauli strings supported on the local domain, so that \(|\mathcal P_{n,k}|\le 4^D\). The generator is then written as
\begin{equation}\label{eq:generatorQITE}
    \boldsymbol{\hat{A}}_{n,k}
    =
    \sum_{I\in\mathcal P_{n,k}}
    a_I^{(n,k)}\boldsymbol{\hat{\sigma}}_I,
\end{equation}
where \(a_I^{(n,k)}\in\mathbb{R}\) are obtained by minimizing the quadratic QITE objective
\begin{equation}
    f(\vec{a}^{(n,k)})
    =
    f_{0}
    +
    \sum_I b_Ia_I
    +
    \sum_{I,J}a_I S_{IJ}a_J.
\end{equation}
Although the locality domain restricts the number of candidate operators, the solution of the standard QITE problem is generally dense, so a large fraction of the \(4^D\) Pauli strings may contribute to the generator. After Trotterizing the corresponding unitary generated by \eqref{eq:generatorQITE}, each nonzero coefficient generally gives rise to a Pauli rotation. The number of active generator terms therefore directly affects the gate count and circuit depth required at each imaginary-time step.

Sparse-QITE modifies the QITE optimization problem by introducing an \(\ell_1\)-regularization term,
\begin{equation}
\label{eq:sparse_qite_cost}
    \vec{a}_{\lambda}^{(n,k)}
    =
    \operatorname*{arg\,min}_{\vec{a}}
    \left[
        f_n(\vec{a})
        +
        \lambda\|\vec{a}\|_1
    \right],
\end{equation}
or, explicitly,
\begin{equation}
\label{eq:sparse_qite_explicit}
    \vec{a}_{\lambda}^{(n,k)}
    =
    \operatorname*{arg\,min}_{\vec{a}}
    \left[
        \sum_{I,J}a_I S_{IJ}a_J
        +
        \sum_I b_Ia_I
        +
        \lambda\sum_I |a_I|
    \right],
\end{equation}
where \(\lambda\geq 0\) controls the strength of the sparsity-inducing penalty. The active support for time step $n$ of the generators resulting from approximating the ITE of all Hamiltonian pieces, is defined as
\begin{equation}
\label{eq:active_support}
    C_n
    =
    \left\{
        I\in\mathcal \cup_{k}P_{n,k}, : a_{\lambda,I}^{(n,k)}\neq 0 \; 
    \right\},
\end{equation}
and its cardinality \(|C_n|\) gives the number of Pauli strings retained at imaginary-time step \(n\)

The locality domain and the regularization play complementary roles. The domain size \(D\) defines the full set of Pauli strings that are available to approximate the local imaginary-time evolution, providing the geometric truncation used in standard QITE. Within this candidate Pauli set, the regularization selects the operators that are dynamically relevant for the current state. Sparse-QITE therefore replaces a generator determined solely by its allowed spatial support with one whose operator content is additionally selected according to its contribution to the QITE approximation.

The regularization parameter controls the trade-off between sparsity and approximation accuracy. For \(\lambda=0\), Eq.~\eqref{eq:sparse_qite_cost} reduces to the standard QITE optimization problem. As \(\lambda\) increases, coefficients that contribute weakly to the reduction of the QITE objective are driven to zero, producing increasingly compact generators, while excessively large values may remove operators that are required to reproduce the imaginary-time step. Above a threshold \(\lambda_{\max}\) the solution vanishes identically, a property used by the adaptive variant of Sec.~\ref{sec:adaptive_sparse_qite}. An appropriate choice of \(\lambda\) can therefore substantially reduce \(|C_n|\) while keeping the Sparse-QITE trajectory close to the standard QITE evolution.

For the basic Sparse-QITE formulation, the Gram matrix \(\boldsymbol{S}\) and vector \(\vec{b}\) are constructed from the same expectation values as in standard QITE. Sparse-QITE therefore initially has the same quantum-measurement requirements as standard QITE, while modifying only the classical post-processing used to determine the generator coefficients.

By eliminating weakly contributing Pauli terms, Sparse-QITE reduces the number of Pauli rotations appearing in the generator and can consequently lower the compiled gate count and circuit depth. This trade-off is particularly favorable because classical optimization is generally less restrictive than coherent quantum execution: reducing circuit depth can mitigate the accumulation of hardware errors and permit longer imaginary-time evolutions or larger QITE domains within a fixed quantum-resource budget. The extent to which these benefits are realized depends on the circuit compilation and hardware model, and is quantified through the compiled-resource analysis presented in Sec.~\ref{sec:experiments}.

As mentioned earlier, we also introduce two extensions of the basic Sparse-QITE formulation. Adaptive Sparse-QITE allows the regularization parameter \(\lambda\) to vary throughout the imaginary-time evolution, while Active-Discovery Sparse-QITE uses the support identified during the evolution to reduce the candidate operator set and, consequently, the measurement requirements. These two variants are described in detail in Secs.~\ref{sec:adaptive_sparse_qite} and~\ref{sec:active_discovery}, respectively.

\subsection{Algorithm implementation}
\label{sec:numerical_implementation}

The optimization problem in Eq.~\eqref{eq:sparse_qite_cost} is convex because the QITE Gram matrix is positive semidefinite and the \(\ell_1\) norm is convex. We solve it using the fast iterative shrinkage-thresholding algorithm (FISTA)~\cite{doi:10.1137/080716542}, which combines gradient steps for the smooth quadratic part of the objective with soft-thresholding operations that promote exact zeros in the coefficient vector. The details of the FISTA implementation used in this work are provided in App.~\ref{app:fista_implementation}, and details on the classical computational cost of Sparse-QITE can be found in App.~\ref{app:computational-cost}.

In practice, the Gram matrix may be singular or ill-conditioned. It is built from the tangent directions
\begin{equation}
\label{eqn:tangdir}
    \ket{e_I^{(n)}}=-i\boldsymbol{\hat{\sigma}}_I\ket{\psi_n}
\end{equation}
generated by the Pauli strings \(I\in\mathcal P_n\), and distinct Pauli strings may act identically, or nearly identically, on the current state. Exact null directions of the Gram matrix then make the coefficient vector nonunique, while near-null directions make it highly sensitive to finite-shot measurement noise and numerical perturbations. To stabilize the coefficient determination, we apply ridge, or Tikhonov, regularization by shifting the symmetrized Gram matrix according to
\begin{equation}
    \Sbar
    \longrightarrow
    \Sbar+\rho I,
\end{equation}
where \(\rho>0\) is a small numerical regularization parameter. At the level of the objective function, this shift is equivalent to adding an \(\ell_2\) penalty, which yields the elastic-net problem
\begin{equation}
\label{eq:elastic_net_qite}
    \vec{a}_{\lambda,\rho}^{(n,k)}
    =
    \operatorname*{arg\,min}_{\vec{a}}
    \left[
        f_n(\vec{a})
        +
        \rho\|\vec{a}\|_2^2
        +
        \lambda\|\vec{a}\|_1
    \right].
\end{equation}
The two penalties play different roles. The \(\ell_1\) term defines the method, since it is the term that produces sparse generators. The \(\ell_2\) term only guarantees a unique and numerically stable solution: with the value \(\rho=10^{-10}\) used in our simulations (Sec.~\ref{sec:numerical_setup}) it acts as a numerical safeguard, and it leaves unchanged the threshold \(\lambda_{\max}\) above which the solution vanishes (App.~\ref{app:adaptive_binary_search}). A two-direction example showing why a pseudoinverse alone does not provide this stability is given in App.~\ref{app:ridge_example}.

\subsection{Error bounds}\label{subsec:error-bounds}
We now state the estimates used to compare Sparse-QITE with standard QITE. The purpose of these bounds is not to prove that the active support selected by Sparse-QITE is optimal. Rather, they quantify the effect of the regularization once the Pauli set used to determine the QITE coefficients has been fixed. We compare the standard QITE coefficient problem of Ref.~\cite{Motta2019} with its ridge-stabilized and elastic-net counterparts, using standard regularization ideas~\cite{tibshirani1996regression,tikhonov1977solutions,ZouHastie2005ElasticNet}.

Fix an imaginary-time step $n$, a Hamiltonian piece $h_k$, and the
corresponding normalized input state. Because the following coefficient,
tangent-vector, and same-input estimates all refer to this single local
update, we suppress both update labels $(n,k)$ for clarity. Thus, throughout
the local analysis,
\begin{equation}
    \mathcal P:=\mathcal P_{n,k},
    \qquad
    \vec a:=\vec a^{(n,k)},
    \qquad
    \vec a_{\lambda,\rho}
    :=
    \vec a_{\lambda,\rho}^{(n,k)},
\end{equation}
and analogously for the other quantities associated with this fixed update.
Regularization labels remain as subscripts. The full physical update labels
are restored when the trajectory analysis begins. Let
\begin{equation}
    M:=|\mathcal P|.
\end{equation}
For the discovery part of the Sparse QITE we will be using $M=4^{D}$ but shall commit to the more general notation; this will also apply to the default and adaptive Sparse QITE approaches which are also to be considered.
All comparisons below use this same Pauli set and the same normalized input
state. In particular, $\vec a$, $\vec a_{\rho}$, $\vec a_{\lambda}$, and
$\vec a_{\lambda,\rho}$ denote ordinary, ridge-only, $\ell_1$-only, and
elastic-net QITE coefficients, respectively.

Define the abbreviated local data
\begin{equation}
    \Sbar:=\Sbar_{\mathcal P_{n,k}}^{(n,k)},
    \qquad
    \vec b:=\vec b^{(n,k)}.
\end{equation}
The quadratic QITE objective is then
\begin{equation}
    f(\vec a)
    =
    f_0
    +
    \vec b^{\,T}\vec a
    +
    \vec a^{\,T}\boldsymbol{S}\vec a,
\end{equation}
where $f_0$ is only the coefficient-independent term. Its gradient is
\(
\nabla f(\vec a)=\vec b+\bigl(\boldsymbol{S}+\boldsymbol{S}^{T}\bigr)\vec a=\vec b+2\Sbar\vec a
\), with $\Sbar$ the symmetrized Gram matrix of Eq.~\eqref{eq:real_gram}, and we define
\begin{equation}
    \vec d:=-\frac12\vec b.
\end{equation}
The ordinary and ridge-stabilized coefficient equations are
\begin{align}
    \Sbar\vec a
    &=
    \vec d,
    \\
    (\Sbar+\rho \boldsymbol{\hat{I}})\vec a_{\rho}
    &=
    \vec d.
\end{align}
For elastic-net Sparse-QITE, let
\begin{align}
    \vec z_{\lambda,\rho}
    &\in
    \partial\|\vec a_{\lambda,\rho}\|_1,
    \\
    \vec d_{\lambda,\rho}
    &:=
    \vec d
    -
    \frac{\lambda}{2}\vec z_{\lambda,\rho}.
\end{align}
Then
\begin{equation}
    (\Sbar+\rho \boldsymbol{\hat{I}})\vec a_{\lambda,\rho}
    =
    \vec d_{\lambda,\rho}.
\end{equation}
For $\rho=0$, this reduces to $\vec a_{\lambda}$ and
$\vec d_{\lambda}$. The factor $1/2$ follows from the unscaled quadratic
convention in Eq.~\eqref{eq:QITE_objective}. The subdifferential
$\partial\|\cdot\|_1$ is defined componentwise in
Sec.~\ref{sec:sparse_regularization}.

Using the abbreviated tangent directions
$\ket{e_I}:=\ket{e_I^{(n)}}$ from Eq.~\eqref{eqn:tangdir}, define
\begin{align}\label{eq:tangent_vectors}
    \ket v
    &:=
    \sum_{I\in\mathcal P}a_I\ket{e_I},
    \\
    \ket{v_{\rho}}
    &:=
    \sum_{I\in\mathcal P}a_{\rho,I}\ket{e_I},
    \\
    \ket{v_{\lambda,\rho}}
    &:=
    \sum_{I\in\mathcal P}a_{\lambda,\rho,I}\ket{e_I}.
\end{align}

\begin{proposition}[Local same-set sparsification bound]
\label{prop:local_same_set_bound}
Let
\begin{equation}
    \kappa:=\lambda_{\min}(\Sbar)\geq0,
    \qquad
    \Sbar\succeq\kappa I.
\end{equation}
Then
\begin{equation}
\label{eq:same_set_coeff_bound}
    \|\vec a_{\lambda,\rho}-\vec a_{\rho}\|_2
    \le
    \frac{\lambda\sqrt M}{2(\kappa+\rho)},
\end{equation}
and
\begin{equation}
\label{eq:same_set_tangent_bound}
    \|\ket{v_{\lambda,\rho}}-\ket{v_{\rho}}\|
    \le
    \frac{\lambda}{2}\sqrt{\frac{M}{\kappa+\rho}}.
\end{equation}
\end{proposition}
The proof is given in App.~\ref{app:local_regularization_bounds}. The result
allows a singular Gram matrix: when $\kappa=0$, the ridge shift results in
$\Sbar+\rho \boldsymbol{\hat{I}}$ positive definite.

\begin{corollary}[Comparison with ordinary QITE]
\label{cor:ordinary_reference_bound}
Let $\Sbar^\dagger$ denote the Moore--Penrose pseudoinverse of the symmetrized Gram
matrix $\Sbar$. Define the ordinary-QITE reference coefficient vector by
\begin{equation}
    \vec a:=\Sbar^\dagger\vec d,
    \qquad
    B:=\|\vec a\|_2.
\end{equation}
This is the minimum-norm representative of the real least-squares solution
of Ref.~\cite{Motta2019} whenever the minimizer is nonunique. Then
\begin{equation}
\label{eq:same_set_ordinary_coeff_bound}
    \|\vec a_{\lambda,\rho}-\vec a\|_2
    \le
    \frac{\frac{\lambda}{2}\sqrt M+\rho B}{\kappa+\rho},
\end{equation}
and
\begin{equation}
\label{eq:same_set_ordinary_reference_bound}
    \|\ket{v_{\lambda,\rho}}-\ket v\|
    \le
    \frac{\frac{\lambda}{2}\sqrt M+\rho B}{\sqrt{\kappa+\rho}}.
\end{equation}
\end{corollary}
The proof is given in App.~\ref{app:local_regularization_bounds}. The two
numerator terms are the $\ell_1$-sparsification contribution and the ridge
bias, respectively. For later local comparisons, define
\begin{equation}
\label{eq:q_lambda_rho_definition}
    q_{\lambda,\rho}
    :=
    \frac{\frac{\lambda}{2}\sqrt M+\rho B}{\sqrt{\kappa+\rho}},
\end{equation}
so that $\|\ket{v_{\lambda,\rho}}-\ket v\|\le q_{\lambda,\rho}$.

Let $F_{\rho}$ denote the exact ridge-QITE one-step map for the fixed local
update; the suppressed labels $(n,k)$ are understood. The trajectory analysis
relies on the stability of this map: for fixed finite $\mathcal P$ and
$\rho>0$, $F_\rho$ is Lipschitz continuous in its input state,
\(\|F_{\rho}(\phi)-F_{\rho}(\varphi)\|\le(1+L_{\rho}\Delta\tau)\|\ket\phi-\ket\varphi\|\),
with a finite constant $L_\rho$ given explicitly in
Eq.~\eqref{eq:L_rho_definition}. This result is stated as
Lemma~\ref{lem:ridge_qite_stability} and proved in
App.~\ref{app:trajectory_stability}.

At this point the local suppression of $(n,k)$ ends. For the trajectory
analysis, we label the ordered local updates by a single index
$r=1,\ldots,R$: update $r$ applies the Hamiltonian piece
$\boldsymbol{\hat h}_{k_r}$ at imaginary-time step $n_r$, and every quantity
attached to it carries the label $r$. Let
\begin{equation}
    \boldsymbol{\mathcal P}
    :=
    \bigl(\mathcal P_{r}\bigr)_{r=1}^{R}
\end{equation}
denote the corresponding sequence of candidate Pauli sets, with
$\mathcal P_r:=\mathcal P_{n_r,k_r}$. In particular, set
\begin{equation}
    M^{(r)}:=|\mathcal P_{r}|.
\end{equation}
Here $\kappa^{(r)}(\phi)$ and $B^{(r)}(\phi)$ denote the
state-dependent versions of the local quantities introduced in
Proposition~\ref{prop:local_same_set_bound} and
Corollary~\ref{cor:ordinary_reference_bound}, with the full update label
restored. Define
\begin{align}
    q_{\lambda,\rho,\mathrm{sp}}^{(r)}(\phi)
    &:=
    \frac{\lambda}{2}
    \sqrt{
    \frac{M^{(r)}}
    {\kappa^{(r)}(\phi)+\rho}
    },
    \\
    q_{\rho,\mathrm{ridge}}^{(r)}(\phi)
    &:=
    \frac{\rho B^{(r)}(\phi)}
    {\sqrt{\kappa^{(r)}(\phi)+\rho}}.
\end{align}
The local contribution entering the trajectory recursion is
\begin{equation}
\label{eq:trajectory_local_q_definition}
\begin{aligned}
    \widetilde q_{\lambda,\rho}^{(r)}
    &:=
    q_{\lambda,\rho,\mathrm{sp}}^{(r)}
    \!\left(\Phi_{r-1}^{\lambda,\rho,\boldsymbol{\mathcal P}}\right)
    \\
    &\quad+
    q_{\rho,\mathrm{ridge}}^{(r)}
    \!\left(\Phi_{r-1}^{\boldsymbol{\mathcal P}}\right).
\end{aligned}
\end{equation}
It is the sum of the Sparse-QITE--to--ridge-QITE and
ridge-QITE--to--ordinary-QITE same-input contributions, evaluated on the two
incoming trajectory states.

\begin{corollary}[Trajectory-level estimate]
\label{cor:trajectory_bound}
Let $\ket{\Phi_R^{\boldsymbol{\mathcal P}}}$ and
$\ket{\Phi_R^{\lambda,\rho,\boldsymbol{\mathcal P}}}$ denote the
ordinary-QITE and Sparse-QITE trajectories generated from the same initial
state. At each ordinary-QITE update, the generator is constructed from the
minimum-norm representative of the ordinary-QITE least-squares solution.
Assume that the combined local first-order remainder is uniformly bounded by
$\Gamma(\Delta\tau)^2$, with a constant $\Gamma$ independent of the update
index. Define
\begin{equation}
    \Lambda_R
    :=
    \Delta\tau
    \sum_{j=1}^{R}L_{\rho}^{(j)}.
\end{equation}
Then
\begin{equation}
\label{eq:qite_sparse_trajectory_bound}
\begin{aligned}
    &\left\|
    \ket{\Phi_R^{\lambda,\rho,\boldsymbol{\mathcal P}}}
    -
    \ket{\Phi_R^{\boldsymbol{\mathcal P}}}
    \right\|
    \\
    &\qquad\le
    e^{\Lambda_R}
    \Delta\tau
    \sum_{r=1}^{R}
    \widetilde q_{\lambda,\rho}^{(r)}
    +
    \mathcal O\!\left(
    e^{\Lambda_R}R(\Delta\tau)^2
    \right).
\end{aligned}
\end{equation}
The $\mathcal O((\Delta\tau)^2)$ contribution is the local first-order
expansion remainder accumulated over the $R$ ordered updates; it is separate
from Trotter and ordinary-QITE replacement errors. The exact discrete bound
and the use of $1+x\le e^x$ are given in
App.~\ref{app:trajectory_stability}.
\end{corollary}
The proof is given in App.~\ref{app:trajectory_stability}. To relate this
estimate to the total error with respect to imaginary-time evolution, let
$\ket{\Psi_R^{\mathrm{ITE}}}$ denote the normalized ITE state after the same
ordered sequence of updates. The triangle inequality gives
\begin{equation}
\label{eq:ite_sparse_total_error}
\begin{aligned}
    \left\|
    \ket{\Phi_R^{\lambda,\rho,\boldsymbol{\mathcal P}}}
    -
    \ket{\Psi_R^{\mathrm{ITE}}}
    \right\|
    &\le
    \left\|
    \ket{\Phi_R^{\lambda,\rho,\boldsymbol{\mathcal P}}}
    -
    \ket{\Phi_R^{\boldsymbol{\mathcal P}}}
    \right\|
    \\
    &\quad+
    \left\|
    \ket{\Phi_R^{\boldsymbol{\mathcal P}}}
    -
    \ket{\Psi_R^{\mathrm{ITE}}}
    \right\|.
\end{aligned}
\end{equation}
The first term is controlled by Corollary~\ref{cor:trajectory_bound}, while
the second is the ordinary QITE-to-ITE error bounded in
Ref.~\cite{Motta2019}. Their combination therefore separates the additional
regularization error of Sparse-QITE from the Trotter and local-domain
replacement errors already present in ordinary QITE.

We emphasize that Corollary~\ref{cor:trajectory_bound} is a worst-case
estimate. The admissible stability constant in Eq.~\eqref{eq:L_rho_definition}
is not optimized and scales as $\rho^{-2}$, so that for the small ridge
parameter used in our numerical experiments ($\rho=10^{-10}$,
Sec.~\ref{sec:numerical_setup}) the prefactor $e^{\Lambda_R}$ is extremely
large. The trajectory estimate should therefore be read as a qualitative
statement: it shows that the additional Sparse-QITE error is controlled by the
accumulated local contributions $\widetilde q_{\lambda,\rho}^{(r)}$. The
quantitatively relevant local errors are those bounded in
Proposition~\ref{prop:local_same_set_bound} and
Corollary~\ref{cor:ordinary_reference_bound}, or controlled directly by the
adaptive criterion of Sec.~\ref{sec:adaptive_sparse_qite}.

\subsection{Adaptive Sparse-QITE}
\label{sec:adaptive_sparse_qite}

The fixed-\(\lambda\) formulation applies the same regularization strength
throughout the evolution, although the Pauli directions required by a local
imaginary-time update depend on the current state. Adaptive Sparse-QITE
instead chooses \(\lambda_n\) independently at each update. We formulate this
choice as a constrained search for the largest regularization strength whose
generator remains within a prescribed local approximation tolerance. The
method thereby promotes the strongest admissible sparsification without
requiring a global value of \(\lambda\) to be tuned in advance. For the rest of this section we will suppress the sub indices relating to a time step or Hamiltonian piece. 

For a candidate \(\lambda\), we solve the elastic-net problem in
Eq.~\eqref{eq:elastic_net_qite} and assess its coefficients using the
unpenalized QITE mismatch
\begin{equation}
    \mathcal R(\lambda)
    :=
    f\!\left(
        \vec a_{\lambda,\rho}
    \right) - f\!\left(
        \vec a_{\mathrm{QITE}}
    \right),
\end{equation}
where the objective, $f$, is defined in Eq.~\eqref{eq:QITE_objective}. One can show that
\(\mathcal R(\lambda)\) determines, up to second order in $\Delta \tau$, an upper bound to the local infidelity between Sparse-QITE and QITE states obtained from the same input state, defined as $F^{\rm{loc}}(\lambda)$ in Eq. \eqref{eq:Floc-def}, in appendix \ref{app:adaptive_binary_search}. For real-states and real Hamiltonians, we get the equality
\begin{equation}
\label{eq:adaptive_residual_infidelity}
    1-F^{\mathrm{loc}}(\lambda)
    =
    \Delta\tau^2\mathcal R(\lambda)
    +O(\Delta\tau^3).
\end{equation}
Moreover, we can relate $\mathcal{R}(\lambda)$ directly to the tangent vector errors defined in Eq.~\eqref{eq:tangent_vectors}; more precisely we have the exact equality
\begin{equation}\label{eq:tangent_vector_error_adaptive}
    \|\ket{v_{\lambda,\rho}}-\ket v\| = \sqrt{\mathcal{R}(\lambda)}.
\end{equation} As a result Adaptive Sparse-QITE controls directly the local first order state error between Sparse-QITE and QITE, and is part of the error that contributes to the total trajectory error derived in Eq.~\eqref{eq:qite_sparse_trajectory_bound} simply by replacing $\widetilde q_{\lambda,\rho}^{(r)}$ with $\sqrt{\mathcal{R}^{(r)}(\lambda)}$.

For
a tolerance \(\varepsilon_{\mathrm{obj}}\), the adaptive rule is
\begin{equation}
\label{eq:adaptive_lambda_rule}
    \widetilde\lambda
    =
    \sup
    \left\{
        \lambda\in[0,\lambda_{\max}]:
        \mathcal R(\lambda)
        \leq
        \varepsilon_{\mathrm{obj}}
    \right\},
\end{equation}
where \(\lambda_{\max}=\max_{I\in\mathcal P}|b_I|\) is the
smallest regularization strength for which the zero coefficient vector is
optimal; a proof is given in App.~\ref{app:adaptive_binary_search}.
This gives a state-dependent search interval without a manually chosen upper
scale. Under the conditions discussed in
App.~\ref{app:adaptive_binary_search}, we locate the boundary in
Eq.~\eqref{eq:adaptive_lambda_rule} by binary search. Compared with scanning
a fixed grid, this resolves the largest admissible \(\lambda\) more
accurately and requires only logarithmically many regularized solves in the
requested search resolution. When a compact generator is sufficient, the
selected value remains large; when more Pauli directions are required, the
search automatically moves toward weaker regularization. The remaining
accuracy hyper-parameter, \(\varepsilon_{\mathrm{obj}}\), directly has a physical interpretation: in the setting of Eq.~\eqref{eq:tangent_vector_error_adaptive}, it bounds the leading local state error between Sparse-QITE and QITE by \(\Delta\tau\sqrt{\varepsilon_{\mathrm{obj}}}\) which in turn directly bounds the state trajectory error. The bracketing, stopping criteria, and additional details and proofs are given in
App.~\ref{app:adaptive_binary_search}.

\subsection{Active-Discovery Sparse-QITE}
\label{sec:active_discovery}

The basic Sparse-QITE formulation reduces the number of active Pauli terms in the generator, but it initially requires the same expectation values as standard QITE because the full candidate operator pool is used to construct the Gram matrix and coefficient vector. We now introduce a second variant, referred to as Active-Discovery Sparse-QITE, which uses the support identified by Sparse-QITE to reduce not only the size of the implemented generator, but also the number of measurements required during the evolution.

At each imaginary-time step \(n\), Sparse-QITE produces an active support defined in Eq.~\eqref{eq:active_support}
containing the Pauli strings selected by the regularized optimization. During an initial discovery stage, we record the operators that become active and construct the cumulative support
\begin{equation}
\label{eq:cumulative_support}
    \mathcal{C}_n
    =
    \bigcup_{m=1}^{n} C_m.
\end{equation}
The set \(\mathcal{C}_n\) therefore contains every Pauli string that has appeared in a Sparse-QITE generator up to step \(n\). If the cumulative support stops growing, this indicates that Sparse-QITE is repeatedly selecting operators that have already been discovered, with few or no new Pauli strings entering the generator.

Once the cumulative support is judged to have stabilized, we freeze the candidate operator pool to the discovered set \(\mathcal{C}_{n_{\mathrm{f}}}\), where \(n_{\mathrm{f}}\) denotes the freezing step. From that point onward, the Gram matrix and coefficient vector are constructed only within this reduced operator space. Consequently, only expectation values associated with the discovered Pauli strings need to be measured during the remaining evolution steps.

The freezing procedure used here is deliberately heuristic. Its purpose is not to provide a rigorous guarantee that no relevant Pauli string will become active at a later time, but rather to provide a simple and practical criterion for exploiting the empirical stabilization of the Sparse-QITE support. The additional overhead is negligible: during the discovery stage, one only needs to record which Pauli strings have appeared and monitor the evolution of their cumulative support.

To quantify its stabilization, we use the Jaccard overlap between the cumulative supports at consecutive steps,
\begin{equation}
\label{eq:cumulative_jaccard}
    J_n
    =
    \frac{
        \left|
            \mathcal{C}_n
            \cap
            \mathcal{C}_{n-1}
        \right|
    }{
        \left|
            \mathcal{C}_n
            \cup
            \mathcal{C}_{n-1}
        \right|
    }.
\end{equation}
Because the cumulative support can only grow, \(\mathcal{C}_{n-1}\subseteq\mathcal{C}_n\), and the Jaccard overlap reduces to
\begin{equation}
    J_n
    =
    \frac{
        |\mathcal{C}_{n-1}|
    }{
        |\mathcal{C}_n|
    }.
\end{equation}
Thus, \(J_n=1\) when no new Pauli strings are discovered at step \(n\), while values below unity indicate that the cumulative support has increased. In practice, we consider the cumulative support to be stable when \(J_n\) remains above a prescribed threshold \(J_{\min}\) for \(L\) consecutive steps. The parameter \(L\), which we refer to as the persistence, controls how long the stability condition must be satisfied before the operator pool is frozen. Larger values of \(L\) provide a more conservative freezing criterion, whereas smaller values allow the measurement reduction to begin earlier.

The measurement reduction obtained with this procedure is expected to be more modest than the reduction in generator size produced directly by sparse regularization. During the initial discovery stage, the full candidate pool must still be measured in order to identify the relevant operators. If the complete evolution contains \(N_{\tau}\) steps and the support is frozen after \(n_{\mathrm{f}}\) steps, a simple proxy for the total measurement cost is
\begin{equation}
\label{eq:active_discovery_measurement_cost}
    \mathcal{M}_{\mathrm{AD}}
    =
    n_{\mathrm{f}}\,\mathcal{M}_{\mathrm{full}}
    +
    \left(
        N_{\tau}-n_{\mathrm{f}}
    \right)
    \mathcal{M}_{\mathrm{frozen}},
\end{equation}
where \(\mathcal{M}_{\mathrm{full}}\) and \(\mathcal{M}_{\mathrm{frozen}}\) denote the per-step measurement costs before and after freezing, respectively. Compared with the standard-QITE cost
\begin{equation}
    \mathcal{M}_{\mathrm{QITE}}
    =
    N_{\tau}\,\mathcal{M}_{\mathrm{full}},
\end{equation}
the achievable saving depends both on how early the support can be frozen and on the size of the discovered pool. Even in the ideal limit in which the post-freezing cost is very small, the initial fraction \(n_{\mathrm{f}}/N_{\tau}\) of the evolution must still be performed using the full pool. Nevertheless, Active-Discovery Sparse-QITE provides a practical way of converting the support discovered during the evolution into a direct reduction of the quantum-measurement cost. The procedure introduces no additional quantum overhead and requires only a small classical cost associated with recording and updating the cumulative support.

\section{Numerical Results}
\label{sec:experiments}

In this section, we numerically assess the accuracy, sparsity, and resource requirements of Sparse-QITE. We first describe the simulation setup and illustrate the behavior of the method through representative ground-state preparation examples, comparing the energy, ground-state fidelity, fidelity with standard QITE, and active generator support throughout the evolution. We then examine the locality structure of the selected Pauli strings and characterize the sparsity--accuracy trade-off as the regularization parameter \(\lambda\) is varied. Next, we investigate how the active support scales with the QITE domain size \(D\) under prescribed fidelity constraints and determine whether the reduction in generator size persists after circuit compilation. The complete analysis is presented for the transverse-field Ising model in both the ordered and disordered phases with periodic boundary conditions, together with selected results for the Heisenberg model, also with periodic boundary conditions, to test the generality of the observed sparsification. Finally, we study the two extensions introduced in Secs.~\ref{sec:adaptive_sparse_qite} and~\ref{sec:active_discovery}: Adaptive Sparse-QITE and Active-Discovery Sparse-QITE.

\begin{figure*}[t]
    \centering
    \includegraphics[width=0.9\linewidth]{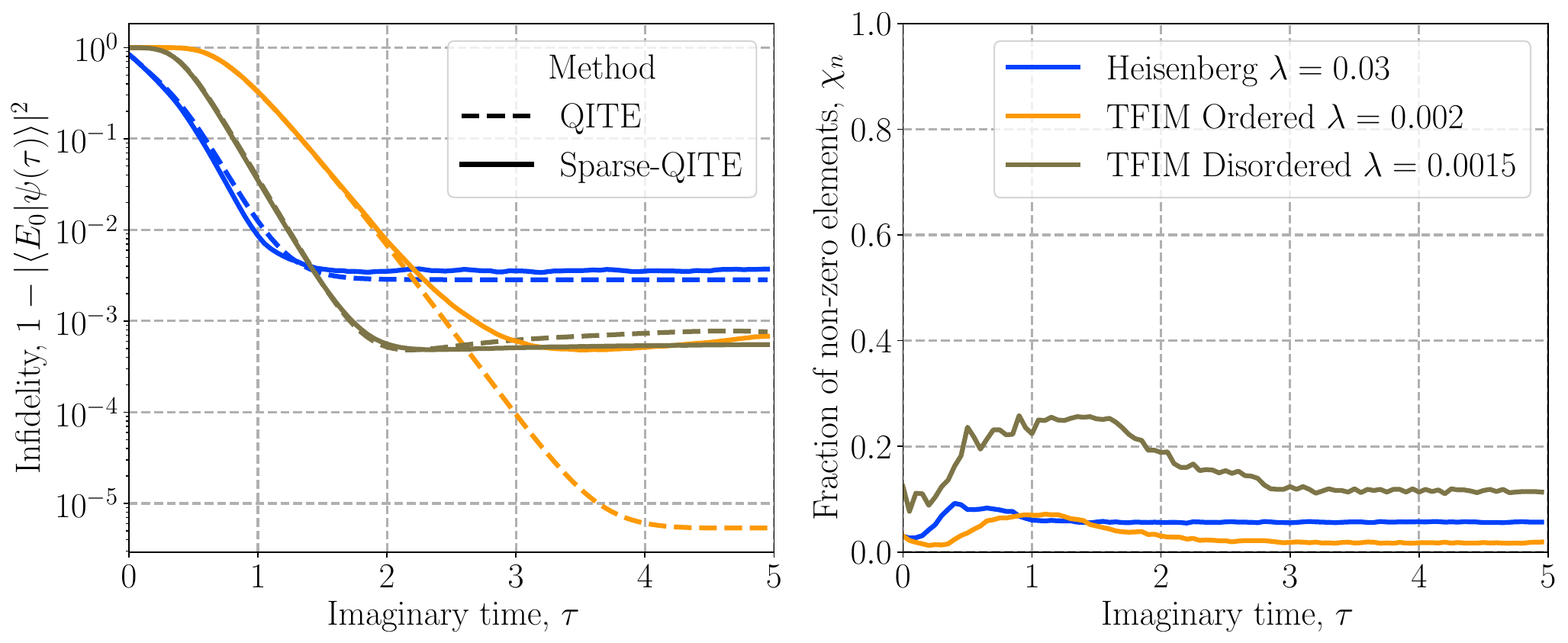}
        \caption{Imaginary-time evolution for the TFIM  (ordered and disordered phase) and the isotropic Heisenberg for a chain of $N=8$ spins. \textbf{(Left)} Ground-state infidelity on a logarithmic scale for each model comparing regular QITE with Sparse-QITE. \textbf{(Right)} Fraction of active Pauli terms $\chi_n$ of Sparse-QITE with respect to QITE. The Heisenberg model results used $D=6$ resulting in $|C_n^{(QITE)}|=7936$, the TFIM ordered-phase results use \(D=6\) with $|C_n^{(QITE)}|=16128$, whereas the disordered-phase results use \(D=4\) with $|C_n^{(QITE)}|=960$.
        In each case, the chosen \(\lambda\) is the largest tested value for which the final-state fidelity with standard QITE is at least \(0.999\) and is given in the legend for each model.}
    \label{fig:tfim_dynamics}
\end{figure*}

\subsection{Numerical Setup}
\label{sec:numerical_setup}

We benchmark Sparse-QITE using the one-dimensional transverse-field Ising model (TFIM) and the isotropic Heisenberg model, both with periodic boundary conditions. The TFIM Hamiltonian is
\begin{equation}
    \hat{H}_{\mathrm{TFIM}}
    =
    -J\sum_{i=1}^{N}
    \hat{Z}_{i}\hat{Z}_{i+1}
    -
    h\sum_{i=1}^{N}
    \hat{X}_{i},
\end{equation}
with \(\hat{Z}_{N+1}:=\hat{Z}_{1}\). We consider \(J=1\) and \(h=0.5\) in the ordered phase, and \(J=0.5\) and \(h=1\) in the disordered phase. The ordered phase is known to have a ground state with longer correlation lengths than the disordered one. As a consequence, to reach good fidelity ground state preparations the ordered phase requires higher sizes of the domain $D$. See \cite{Motta2019} for the study relation correlation lengths and $D$, and \cite{ACQ} for the comparative performance of choosing different $D$ for this Hamiltonian. 

We also consider the one-dimensional antiferromagnetic isotropic Heisenberg Hamiltonian
\begin{equation}
    \hat{H}_{\mathrm{Heis}}
    =
    J
    \sum_{i=1}^{N}
    \left(
        \hat{X}_{i}\hat{X}_{i+1}
        +
        \hat{Y}_{i}\hat{Y}_{i+1}
        +
        \hat{Z}_{i}\hat{Z}_{i+1}
    \right),
\end{equation}
with periodic boundary conditions and \(J=1\).

Unless stated otherwise, all simulations use an imaginary-time step of \(\Delta\tau=0.05\) and are evolved for \(N_{\tau}=100\) steps. The system size \(N\), QITE domain size \(D\), and regularization parameter \(\lambda\) are specified separately for each experiment.

For the transverse-field Ising model, the initial state is chosen as the computational-basis product state
\begin{equation}
    \ket{\psi_0}
    =
    \ket{0}^{\otimes N}.
\end{equation}
For the antiferromagnetic Heisenberg model, the fully polarized state \(\ket{0}^{\otimes N}\) is an excited eigenstate and would therefore remain unchanged under imaginary-time evolution up to normalization. We instead use the Néel state
\begin{equation}
    \ket{\psi_0}
    =
    \ket{0101\cdots},
\end{equation}
which has nonzero overlap with the antiferromagnetic ground state.
The exact ground states and ground-state energies used as references are obtained through exact diagonalization.

To stabilize the QITE Gram matrix, we use a ridge parameter \(\rho=10^{-10}\) throughout the numerical experiments. A Pauli string is considered inactive whenever the absolute value of its associated coefficient is smaller than \(10^{-14}\). Further details of the FISTA implementation, including the convergence criteria, step-size selection, and debiasing procedure, are provided in App.~\ref{app:fista_implementation}.

The implementations of standard QITE and Sparse-QITE, together with the data and scripts used to generate the numerical results, will be made publicly available upon publication of this work.

\begin{figure*}[t]
    \centering
    \includegraphics[width=\linewidth]{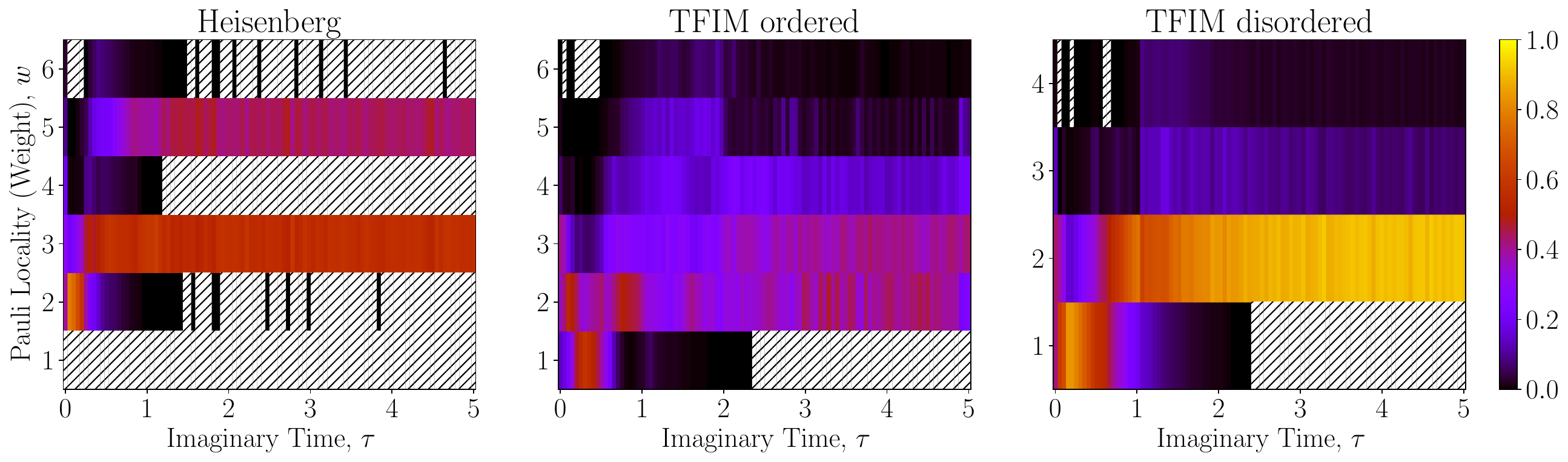}
    \caption{
    Normalized active support by Pauli locality $w$ of Sparse-QITE for the models explored in Fig.~\ref{fig:tfim_dynamics}. Each horizontal entry gives the fraction of active Pauli strings at a given locality, normalized by the total number of available strings at that locality. The white hatched area denotes exactly zero active terms.}
    \label{fig:tfim_locality_heatmaps}
\end{figure*}

\subsection{Sparse-QITE}
\label{sec:baseline_results}

We first illustrate the behavior of Sparse-QITE through representative imaginary-time evolutions and compare it directly with standard QITE. For each benchmark, we monitor the energy expectation value,
\begin{equation}\label{eq:exp_energy}
    E(\tau)
    =
    \bra{\phi(\tau)}
    \hat{H}
    \ket{\phi(\tau)},
\end{equation}
the fidelity with the exact ground state,
\begin{equation}
    F_{\mathrm{GS}}(\tau)
    =
    \left|
        \langle E_0 | \phi(\tau)\rangle
    \right|^2,
\end{equation}
and the fidelity between the Sparse-QITE and standard QITE states,
\begin{equation}
    F_{\mathrm{QITE}}(\tau)
    =
    \left|
        \left\langle
            \phi_{\mathrm{QITE}}(\tau)
            \middle|
            \phi_{\mathrm{Sparse}}(\tau)
        \right\rangle
    \right|^2.
\end{equation}
We additionally record the total number of active Pauli coefficients $|C_n|$ as defined in Eq.~\eqref{eq:active_support} appearing in the local generators at each imaginary-time step. 
We can define the fraction of active terms of Sparse-QITE with respect to the number of active terms in QITE as
\begin{equation}
    \chi_n = \frac{|C_n^{(SpQITE)}|}{|C_n^{(QITE)}|}~,
\end{equation}
where the superindex indicates which method was used to compute the evolution. 
Together, these quantities help determine whether Sparse-QITE preserves the physical evolution produced by standard QITE while reducing the size of its generators.

The regularization strength used in each representative example is selected from the grid search discussed in Sec.~\ref{sec:lambda_sensitivity}. More precisely, for each model and domain size, we choose the largest tested value of \(\lambda\) for which the final-state fidelity satisfies
\begin{equation}\label{eq:criterion_lambda}
    F_{\mathrm{QITE}}(\tau_f)\geq 0.999.
\end{equation}
Within the tested grid, this criterion selects the strongest admissible regularization while maintaining a prescribed accuracy with respect to standard QITE.

Figure~\ref{fig:tfim_dynamics} compares fidelity dynamics of QITE and Sparse-QITE for different models, as well as the fraction of remaining terms in Sparse-QITE for the criterion \eqref{eq:criterion_lambda} that selects the regularization parameter $\lambda$. No curves for the evolution of the energy \eqref{eq:exp_energy} are shown since for this high fidelity criterion they are indistinguishable between methods. 

The ground-state infidelity is shown on a logarithmic scale to provide a more sensitive comparison. In the cases where the final fidelity reaches values similar to or above the one set in Eq.~\eqref{eq:criterion_lambda}, both QITE and Sparse-QITE display very similar trends. Only in the case where the increased domain size of the ordered phase of the TFIM allows to reach more precise fidelities we see the departure of the Sparse-QITE evolution reaching a value of the order set in Eq.~\eqref{eq:criterion_lambda}. These observations indicate that the fidelity impact of the regularization for the chosen $\lambda$ is minimal.

Figure~\ref{fig:tfim_dynamics} also explores the relative size of the active support used by Sparse-QITE. In general the reduction of Pauli generators is substantially reduced, reaching as low as 2\% usage in the best case scenario. It is worth noting that in the disordered phase of the TFIM, where $D=4$, the reduction is less pronounced in relative terms, since with $D=4$ good fidelities are already reached, there is less wiggle-room to remove Pauli generators. When moving to more complex ground states (Heisenberg and TFIM ordered), which require bigger $D$ to capture correlations the regularization excels at removing unnecessary generators, and hence has reduced the active fraction $\chi_n$. More detailed numerical values are reported in App.~\ref{app:baseline_data} for the TFIM model and in App.~\ref{app:heisenberg_results} for the Heisenberg model.

\begin{figure*}[t]
    \centering
    \includegraphics[width=0.7\linewidth]{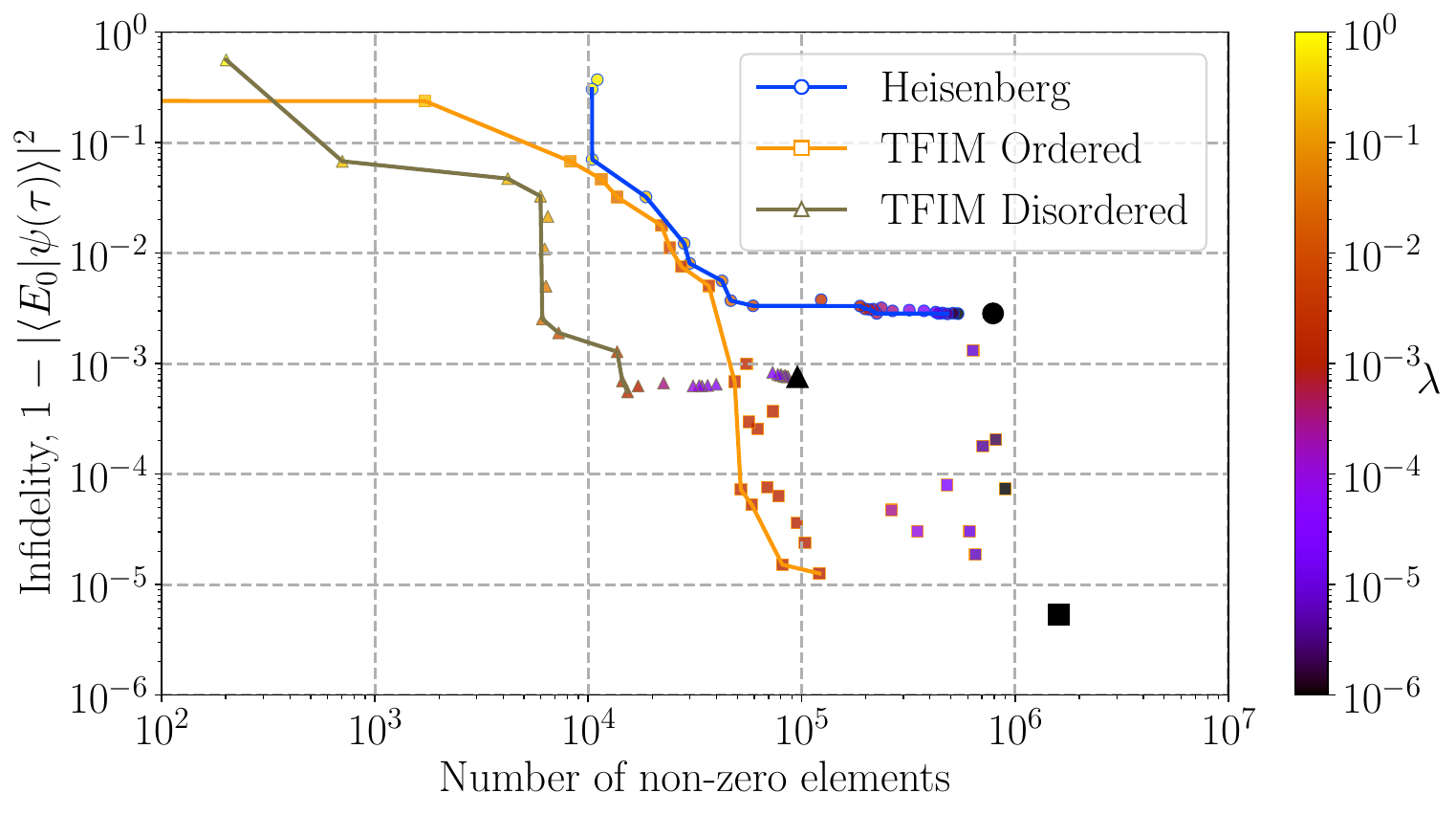}
    \caption{
    Sparsity--accuracy trade-off for Sparse-QITE of the models studied in Fig.~\ref{fig:tfim_dynamics} sweeping the regularization parameter $\lambda$, where each point corresponds to a different value. The horizontal axis gives the total number of active Pauli rotations accumulated over the trajectory of 100 imaginary time steps, while the vertical axis shows the final ground-state infidelity. Solid lines represent the Pareto front containing the points that minimizes infidelity and active Pauli terms. Solid black markers denote standard QITE costs and final infidelities.}
    \label{fig:tfim_pareto}
\end{figure*}

\subsubsection{Pauli Weights Sparsity}
To investigate which classes of Pauli strings are removed, we analyze the active support according to the Pauli locality. For a given locality \(w\), we define the normalized active fraction as the number of active Pauli strings of locality \(w\) divided by the total number of Pauli strings of the same locality in the full operator set. This normalization allows the different locality sectors to be compared despite containing different numbers of operators.

For standard QITE essentially all available Pauli strings become active at every locality\footnote{Apart from the first update where the simple initial states considered only activate few Pauli generators, but after that step all become active.}.
Figure~\ref{fig:tfim_locality_heatmaps} therefore focuses on the locality-resolved support selected by Sparse-QITE, that is looking at the fraction of active Pauli coefficients with the same $w$. It can be observed that the fraction of retained operators generally decreases as the Pauli locality increases. Low-locality (and lower cardinality) sectors contain the largest active fractions during the initial part of the evolution, when the state changes most rapidly, whereas only a small fraction of the higher-locality strings remains active. Interestingly, after this initial transient, all one-body Pauli strings are completely removed and remain inactive for the rest of the evolution. In particular, for the Heisenberg model they are not present at all. More generally, the active support evolves nontrivially with imaginary time: it initially expands as the state departs from the initial configuration and subsequently contracts as the evolution approaches convergence, a fact that can be observed in the $\chi_n$ plot of Fig.~\ref{fig:tfim_dynamics} and motivates the method described in Sec.~\ref{sec:active_discovery}.


This separation between locality sectors is especially pronounced in the TFIM ordered phase. Although a domain of size \(D=6\) is required for standard QITE to accurately reproduce the ground-state preparation, only a small fraction of the Pauli strings at the largest allowed localities survives the regularization. Sparse-QITE therefore benefits from the expressive power of the larger domain without retaining the complete exponentially large set of Pauli strings in every generator.

The heatmaps also show that the sparsification is not equivalent to simply replacing the original domain \(D\) with a smaller fixed domain. Sparse-QITE retains contributions across several locality sectors, including some higher-locality strings, but selects only a small and time-dependent subset of the available operators. The method thus constructs a dynamically adapted compact generator within the original QITE domain.

\subsubsection{Sparsity--Accuracy Trade-off}
\label{sec:lambda_sensitivity}

We next investigate the sensitivity of Sparse-QITE to the regularization parameter \(\lambda\). Increasing \(\lambda\) strengthens the sparsity penalty and generally reduces the number of active Pauli terms, but also increases the deviation from the target imaginary-time evolution. Conversely, smaller values of \(\lambda\) recover increasingly accurate trajectories at the cost of retaining more terms. The regularization parameter therefore provides direct control over the trade-off between generator compactness and physical accuracy. To systematically characterize this trade-off, we first sample the regularization strength on the logarithmic grid
\[
\lambda \in
\left\{
10^{-5},\,3\times10^{-5},\,
\ldots,\,
10^{-1},\,3\times10^{-1},\,
1
\right\}.
\]
and subsequently introduce finer values in the regions where the balance between fidelity and sparsity changes most rapidly.


\begin{figure*}[t]
    \centering
    \includegraphics[width=\linewidth]{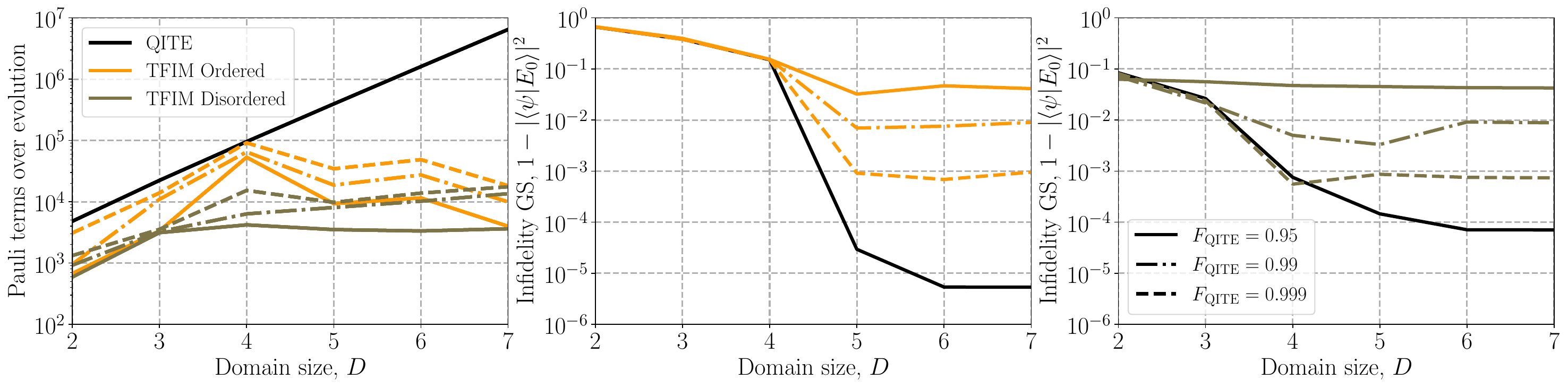}
    \caption{
    Scaling of Sparse-QITE with the QITE domain size \(D\) for the \(N=8\) TFIM.  \textbf{(Left)} Total number of active Pauli terms accumulated over the full imaginary-time trajectory. \textbf{(Center)} Final ground-state infidelity for ordered phase. \textbf{(Right)} Final ground-state infidelity for disordered phase. For each fidelity threshold with respect to standard QITE, the largest tested regularization strength \(\lambda\) satisfying the threshold is selected. Standard-QITE results are included in solid black lines for comparison. Both the active-term counts and ground-state infidelities are displayed on logarithmic scales.
    }
    \label{fig:tfim_scalability}
\end{figure*}


Figure~\ref{fig:tfim_pareto} shows this trade-off for the same models as before. Infidelity and active Pauli terms during the whole evolution are recorded for different values on $\lambda$. In all cases, the variations in \(\lambda\) traces the anticipated sparsity--accuracy trade-off. Moreover, the Pareto curves reveal some interesting features. 
First, for the disordered phase of the TFIM model, it can be seen that, at some point, decreasing $\lambda$ has little impact on the size of the active set, but has a significant improvement in fidelity. This fact indicates that the reduction in active terms and fidelity is not linear, and that dynamics are relevant to the regularization scheme. This observation further motivates the adaptive regularization methodology presented in Sec.~\ref{sec:adaptive_sparse_qite}. 
Second, depending on the ability for QITE to prepare the ground state a minimum $\lambda$ can saturate the maximum achievable fidelity. This fact is observed in the Heisenberg which for approximately $\lambda < 10^{-2}$ the gains in fidelity are negligible, while the number of active terms is increased. This fact is also oberved for the tail end of the TFIM disordered phase. For the ordered phase of the TFIM, whose domain size $D=6$ is large, QITE can reach higher fidelities and this fact is not as prononced but still present. 
Finally, the fidelities reached by QITE can be achieved by Sparse-QITE with up to two orders of magnitude less Pauli generators if the appropiate $\lambda$ is chosen.

These results therefore demonstrate two complementary effects of the regularization. First, \(\lambda\) controls a systematic trade-off between generator sparsity and final-state accuracy. Second, Sparse-QITE can eliminate many terms that are not required to attain the accuracy reached by standard QITE, yielding substantially more compact generators even in the high-accuracy regime. Representative numerical values for the different sparsity and accuracy regimes are reported in Table~\ref{tab:tfim_lambda_sensitivity} in App.~\ref{app:tfim_lambda_sensitivity}.

\subsubsection{Scaling with the QITE Domain Size}
\label{sec:scalability}

A central limitation of standard QITE is that the local Pauli operator set grows exponentially with the domain size \(D\), containing \(4^D\) Pauli strings before symmetry reductions. Although increasing \(D\) can improve the approximation of the imaginary-time evolution, it also rapidly increases the number of coefficients that must be determined and the number of terms in the corresponding unitary generator. Since each active Pauli term translates into a quantum operation, this growth directly increases the circuit gate count and, in general, the circuit depth. The purpose of the following experiment is therefore to determine whether Sparse-QITE can exploit the enlarged operator set while retaining a substantially smaller active subset. In particular, we investigate whether the number of active Pauli terms required to reproduce QITE at a prescribed accuracy grows substantially more slowly with \(D\) than the size of the full QITE generator.

For each \(D=2,\ldots,7\), we run Sparse-QITE over the same grid of regularization strengths \(\lambda\) considered in Sec.~\ref{sec:lambda_sensitivity}. 
This grid is extended by one order of magnitude, down to \(\lambda=10^{-6}\), for \(D=6\) and \(7\). For a prescribed final-state fidelity threshold with respect to standard QITE, we select the largest value of \(\lambda\) whose final state satisfies that threshold. In this way, we retain the strongest regularization compatible with the required agreement with the QITE evolution. We consider thresholds of \(0.95\), \(0.99\) and \(0.999\).
A comparison for the accumulated active Pauli terms and ground state fidelity is studied in Fig.~\ref{fig:tfim_scalability}. This comparison makes it possible to assess both the reduction in generator size and the extent to which the remaining ground-state error originates from sparsification rather than from the intrinsic accuracy of QITE at each domain size.


The results in Fig.~\ref{fig:tfim_scalability} show a clear trade-off between sparsity and accuracy across the different domain sizes as already studied in the previous section. It can be seen that relaxing the required agreement with standard QITE leads to smaller generators, whereas stricter fidelity thresholds retain more terms and reproduce the QITE evolution more closely. At the same time, substantial compression remains possible even under stringent fidelity requirements, in some cases yielding results that are nearly indistinguishable from standard QITE. 


\begin{figure*}[t]
    \centering
    \includegraphics[height=4.7cm]{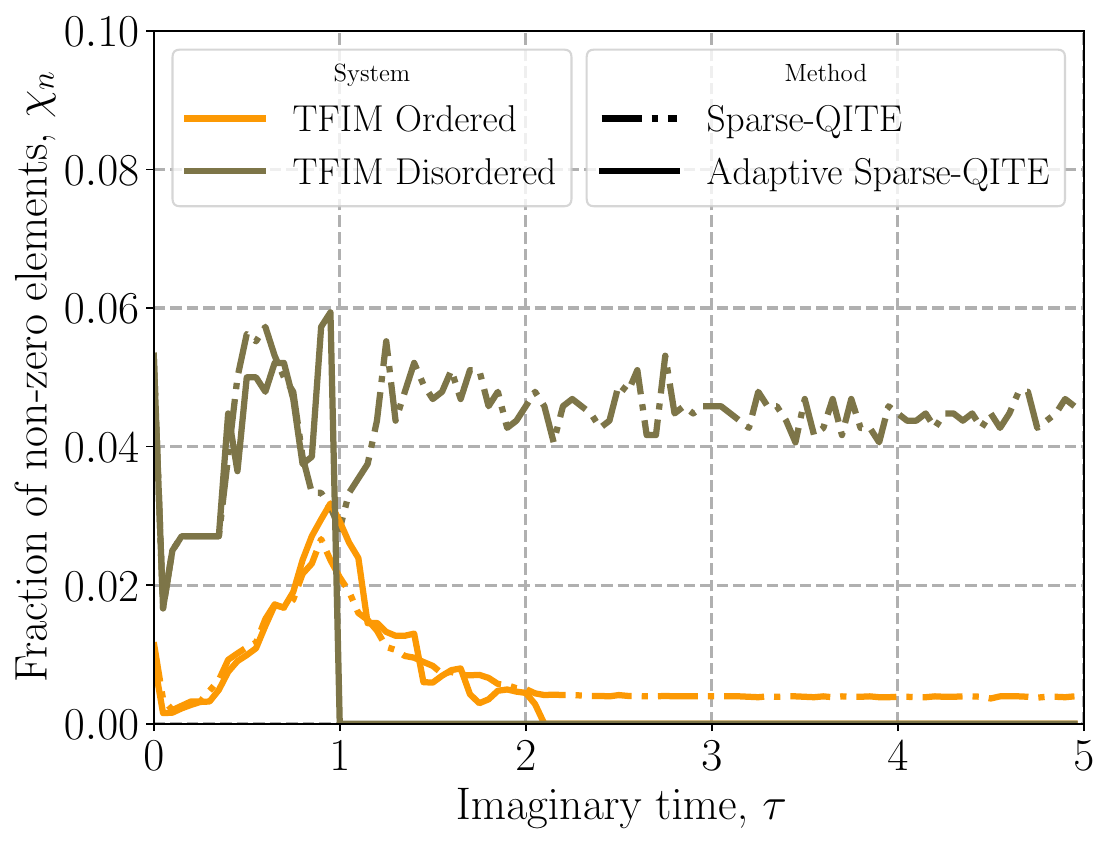}
    \hfill
    \includegraphics[height=4.7cm]{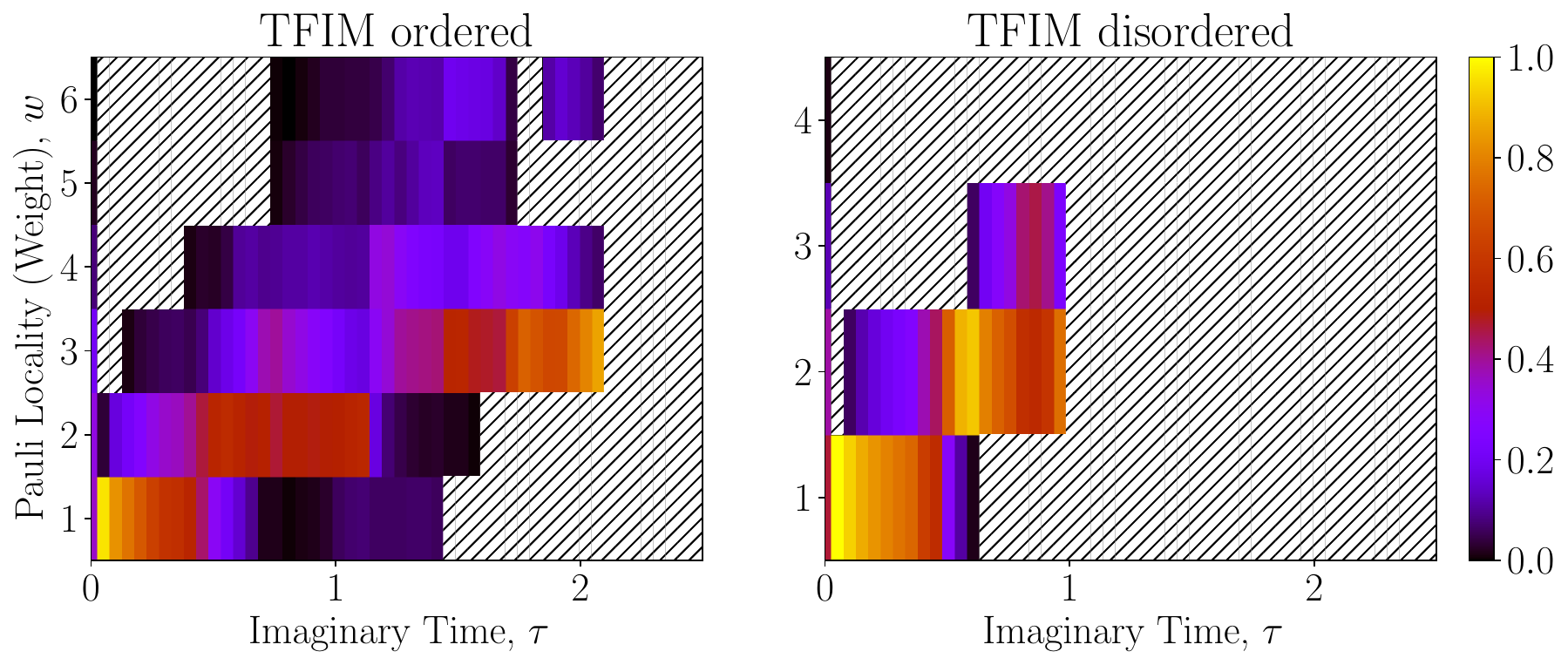}
    \caption{Matched-accuracy comparison between fixed and Adaptive Sparse-QITE for the \(N=8\) TFIM for the same phases as previous figures. The fixed and adaptive hyperparameters were chosen to give minimum fidelities with the standard-QITE trajectory of approximately \(0.95\). In particular, for the disordered phase $\lambda = 0.25$ and $\varepsilon_\mathrm{obj} =0.113$, while for the ordered phase $\lambda=0.057$ and $\varepsilon_\mathrm{obj}=0.0224$. \textbf{(Left)} Fraction of total active Pauli terms with respect to QITE for both methods. \textbf{(Center)} Heatmap showing the fraction of active Pauli terms of the adaptive method per Pauli weight $w$ as function of imaginary time for the ordered phase and \textbf{(Right)} disordered phase. The white hatched area denotes exactly zero active terms.}
    \label{fig:adaptive_vs_fixed_dynamics}
\end{figure*}


While the generator size of QITE is exponential with $D$ and unbounded, Sparse-QITE shows a substantial reduction, and even for the case of the ordered TFIM, a reduction with increasing $D$. This fact is only observed in the ordered phase, where the longer correlation lengths of the ground state make considering higher $D$ terms more relevant to reach the ground state. The sparsification at higher $D$ allows to remove irrelevant terms that only contribute marginally to better fidelities, hence, with fewer terms (with bigger domains) reach better results. On the other hand, even though the disordered phase has in general fewer active terms, it does not benefit by considering higher $D$ generators. The Heisenberg model is not shown in Fig.~\ref{fig:tfim_scalability} for simplicity, but displays similar features as the TFIM disordered phase. We can conclude from these results, that Sparse-QITE for highly correlated ground state problems benefits from considering higher locality generators, yielding a reduced number of active Pauli terms without any detriment to the reached fidelity.

The present results indicate that Sparse-QITE may allow larger domains to be explored without inheriting the full growth of the standard-QITE generator. 
Sparse-QITE may therefore be especially useful when the domain required for convergence is not known in advance, since it can make the exploration of larger values of \(D\) substantially more resource-efficient in terms of circuit complexity.
The complete numerical values for the selected regularization strengths, active-term counts, and final ground-state infidelities are reported in App.~\ref{app:baseline_data}.

\subsubsection{Resource cost}

To assess how the reduction in active Pauli terms translates into actual circuit-resource savings, it is needed to compile the unitary transformations generated by standard QITE and Sparse-QITE into elementary one- and two-qubit gates. The exponentials of the Pauli generators are usually implemented using a second-order symmetric Suzuki decomposition and subsequently transpiled in Qiskit to the gate set \(\{\mathrm{u3},\mathrm{cx}\}\). This allows for a direct comparison of the reduction in generator size directly with the resulting gate counts and circuit depths after compilation.

We perform this compilation analysis for the same set of simulations used in the scalability study, covering the different domain sizes and fidelity thresholds considered above in 
Sec.~\ref{sec:scalability}. The detailed construction of the Suzuki-decomposed unitaries and the transpilation procedure are described in App.~\ref{app:compilation_scalability}, together with the numerical analysis of the resulting reduction in gate counts.  
The Suzuki decomposition and subsequent transpilation can introduce an additional approximation error, but, since this source of error is of second order in the time step, the compiled states retain essentially the same ground-state accuracy as the corresponding uncompiled evolutions. This fact indicates that the large circuit-resource reductions are not accompanied by a significant degradation of the prepared state.

\subsection{Adaptive Sparse-QITE}
\label{sec:adaptive_results}

To compare the Adaptive Sparse-QITE proposal presented in Sec.~\ref{sec:adaptive_sparse_qite} with the initial fixed-regularization method we set to compare evolution at a matched accuracy with respect to standard QITE of both methods.  The comparison is based on the minimum fidelity with QITE over the complete saved trajectory, rather than only on the final state. We fixed a target fidelity with QITE and probed different $\lambda$ values until the fidelity reached the desired value along the evolution. Similarly, the same procedure was performed to set the hyperparameter $\varepsilon_\mathrm{obj}$, which
constrains the unregularized QITE mismatch locally through Eq.~\eqref{eq:adaptive_residual_infidelity}.

\begin{figure*}[t]
    \centering
    \includegraphics[width=\linewidth]{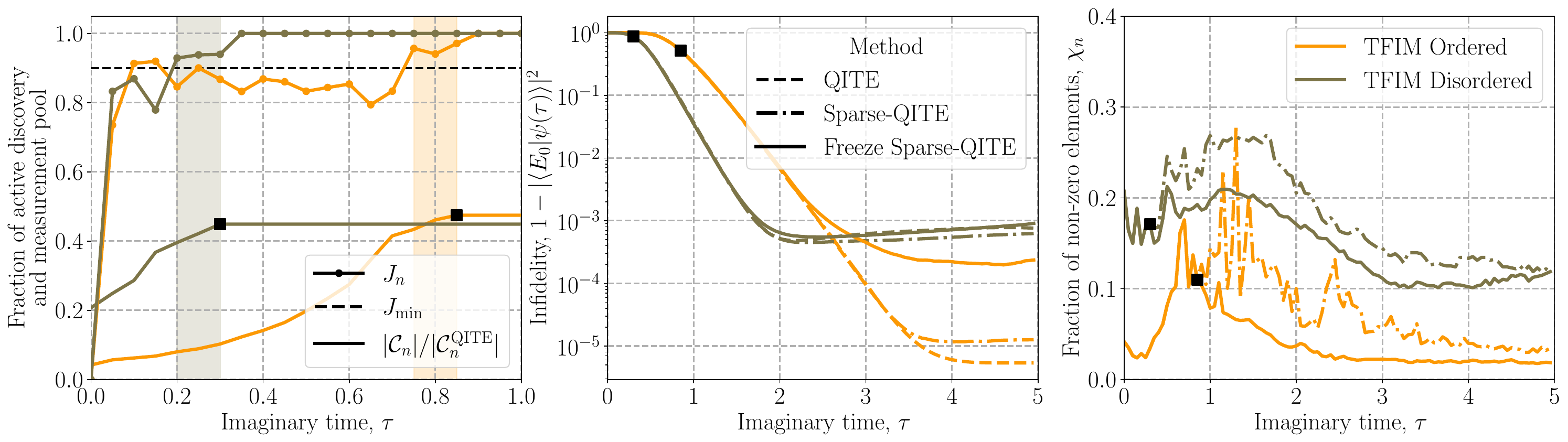}
    \caption{
    Representative Active-Discovery Sparse-QITE evolutions for the TFIM, with the same values as previous figures, using a fixed \(\lambda=0.001\), cumulative-support threshold \(J_{\min}=0.9\), and persistence \(L=3\).
    \textbf{(Left)} Jaccard overlap $J_n$ defined in Eq.~\eqref{eq:cumulative_jaccard} and the cumulative measurement pool fraction with respect to the QITE one $|\mathcal{C}_n|/|\mathcal{C}_n^\mathrm{QITE}|$. The shaded region captures the first $L=3$ steps in which $J_n>J_\mathrm{min}$, and the black square the point from which the pool of Pauli measurements is frozen. 
    \textbf{(Center)} Infidelity with the ground state for QITE, Sparse-QITE and Freeze Sparse-QITE.
    \textbf{(Right)} Fraction of non-zero Pauli generators with respect to QITE for both Sparse-QITE and Active-Discovery QITE.
    }
    \label{fig:active_discovery_tfim}
\end{figure*}

Figure~\ref{fig:adaptive_vs_fixed_dynamics} shows that the adaptive freedom produces additional compression beyond what can be obtained with one fixed value of \(\lambda\).  
Both models studied there present a sharp change in behavior, that occurs at different times, after which the selected generator contains no active coefficient, and hence every subsequent Sparse-QITE unitary is the identity.
Operationally, the adaptive criterion has determined that the zero update already lies within the permitted local QITE-mismatch budget.  The approximate state therefore remains fixed for the rest of the evolution.  In this example, the deviation from the QITE trajectory is established during the early updates; after these sharp supressions of Pauli weights, the remaining QITE corrections are sufficiently small that no further nontrivial unitary is required by the prescribed local tolerance.  This behaviour of the adaptive algorithm serves as a stopping-like criterion which emerges directly from the local error criterion and is not encoded in fixed-\(\lambda\) Sparse-QITE.


The heatmaps in Fig.~\ref{fig:adaptive_vs_fixed_dynamics} show the difference in the active support weight over time more clearly. Notice that after some time all Pauli generators are suppressed. 
In the ordered phase of the TFIM, which contains longer range correlations, Pauli terms of all allowed weights are activated, but longer range generators are progressively activated, in correspondence to the entanglement generation with imaginary time. This is in contrast to the fixed $\lambda$ method used before. This adaptive sparsification selects just the required generators that are most relevant for the ground state preparation that here is shown to be time dependent.  In the disordered phase, longer range generators of $w=4$ are totally suppressed, indicating that this phase possesses correlations of shorter length.


The adaptive method provides two advantages simultaneously. It replaces the global regularization parameter $\lambda$ by $\varepsilon_{\mathrm{obj}}$ which is related to the local infidelity with QITE that also enters the error bounds discussed in Sec \ref{subsec:error-bounds} and \ref{sec:adaptive_sparse_qite}. Moreover, it can exploit the changing structure of the QITE update to achieve additional sparsity at matched trajectory accuracy.

\subsection{Active-Discovery Sparse-QITE}
\label{sec:active_discovery_results}

We now illustrate the behavior of Active-Discovery Sparse-QITE for the same TFIM benchmarks considered above. In this case, we fix the regularization strength to \(\lambda=0.001\) and use the heuristic freezing criterion introduced in Sec.~\ref{sec:active_discovery}, with cumulative-support Jaccard threshold \(J_{\min}=0.9\) and persistence \(L=3\). Thus, the candidate operator set is frozen once the cumulative support overlap \(J_n\) remains above \(0.9\) for three consecutive imaginary-time steps. This means that no substantial enlargement of the discovered support is observed over that interval, indicating that the relevant Pauli strings have largely been identified.

Our goal here is not to provide a full exhaustive investigation of this variant, but rather to demonstrate its potential as a practical extension of Sparse-QITE. In particular, we aim to show that the cumulative active support can stabilize sufficiently early in the evolution so that the operator set may be frozen while still retaining good agreement with both standard QITE and the exact ground state. When this occurs, the number of expectation values that must be measured during the remaining evolution is reduced, thereby turning the support discovered by Sparse-QITE into a direct measurement saving.

Figure~\ref{fig:active_discovery_tfim} shows representative Active-Discovery Sparse-QITE trajectories for the ordered and disordered phases of the \(N=8\) TFIM. 
In both phases, the cumulative support $J_n$ grows rapidly during an initial discovery stage and subsequently enters a regime of much slower variation, allowing the operator set to be frozen while preserving an accurate evolution. 
Here, we quantify the measurement reduction through the cumulative number of Pauli-string expectation values evaluated over the complete imaginary-time evolution. The measurement pool reduction here is more modest than the sparsification at each steps, displaying reduction of lower than 50\%. This reduction however has a negligible impact on fidelity as can be seen in the central panel of Fig.~\ref{fig:active_discovery_tfim}. This reduced accessible set of Pauli measurements comes with a further advantage in sparsification of Pauli generators. 



These representative results show that the support identified during the early stages of Sparse-QITE can be reused throughout the remaining evolution without significantly degrading the prepared state. The resulting measurement savings are necessarily more modest than the reduction in generator size, since the full operator set must still be measured during the initial discovery stage. Nevertheless, the results demonstrate that Active-Discovery Sparse-QITE can convert the sparsity identified by the regularized evolution into a direct reduction in the number of Pauli expectation values that need to be evaluated and, consequently, into a reduction in the number of shots required on the quantum computer to estimate those quantities.

\section{Conclusions}
\label{sec:conclusions}

We introduced Sparse-QITE, a regularized formulation of quantum
imaginary-time evolution that promotes compact local generators by selecting
only a subset of the Pauli strings available within the QITE domain. The
generator coefficients are obtained through elastic-net regularization: the
$\ell_1$ contribution induces sparsity, while a small $\ell_2$ term stabilizes
the solution of the underlying QITE linear system. We also derived bounds on
the additional approximation error introduced by regularization, connecting
the degree of compression to the deviation from standard QITE.

Our numerical results for the transverse-field Ising and antiferromagnetic
Heisenberg models demonstrate a controllable sparsity--accuracy trade-off.
More importantly, substantial compression persists in regimes where Sparse-QITE closely reproduces the standard-QITE trajectory. In several cases,
ground-state accuracies comparable to QITE are obtained using only a small
fraction of the full generator. This shows that the observed reduction is not
solely obtained by accepting a large accuracy loss, but also reflects
substantial redundancy in the standard-QITE operator representation.

This separation becomes increasingly pronounced as the QITE domain size $D$
is enlarged. Whereas the full local Pauli pool grows as $4^D$, the number of
active terms selected by Sparse-QITE grows much more slowly over the domain
sizes studied and, in some regimes, appears to approximately saturate once the
prescribed accuracy has been reached. Although the accessible system and
domain sizes are insufficient to establish an asymptotic scaling law, the
results provide evidence that accurate QITE trajectories can admit
representations substantially more compact than the full local operator
basis. This behavior may be particularly useful when the domain required for convergence is not known in advance, since larger domains can be explored
without necessarily retaining their full operator spaces.

The compression also survives circuit construction. After Suzuki
decomposition and transpilation, the reduced generators yield substantial
decreases in one- and two-qubit gate counts and circuit depth, with the
largest savings occurring at the largest domains considered. Sparse-QITE
therefore reduces circuit resources rather than only the formal
number of generator coefficients. The two extensions introduced here exploit
this structure further. Adaptive Sparse-QITE selects the regularization
strength independently at each update according to a locally interpretable
approximation tolerance, providing additional compression at matched
trajectory accuracy. Active-Discovery Sparse-QITE instead restricts later
updates to the support identified during an initial discovery stage, thereby
converting the learned sparsity into a reduction of the required expectation values. Under independent, equal-precision estimation, this reduction also
corresponds to a lower shot requirement.

Several questions remain open. A central theoretical problem is to determine
how the compressibility of QITE generators depends on locality, correlation
length, symmetry, entanglement, and proximity to criticality, and whether the
favorable dependence on $D$ persists for larger systems and more strongly
correlated models. The active-discovery strategy could also be generalized to
allow operators to leave and re-enter the candidate pool, and combined with
measurement-grouping or classical-shadow protocols. Finally, testing
Sparse-QITE under realistic noise will determine whether its lower gate
counts and depths translate into improved hardware performance. More broadly,
the same regularization principle may apply to other quantum algorithms that
determine generators or circuit parameters through optimization over large
operator sets, for example, in variational quantum algorithms.

\section*{Acknowledgements}

This work was supported by funding from Horizon Europe EU projects MSCA-SE CaLIGOLA, Project
ID: 101086123, and MSCA-DN CaLiForNIA, Project ID: 101119552. RGL is funded by grant CIACIF/2021/136 from Generalitat Valenciana.

\section*{Code Availability}

Code that implements the proposed algorithm will be made available in the published version of this work.

\bibliography{biblio}
\newpage

\onecolumngrid
\appendix
\section{FISTA implementation}
\label{app:fista_implementation}

This appendix summarizes how FISTA is solved in
our implementation. For each time step and local Hamiltonian piece, the
elastic-net problem in Eq.~\eqref{eq:elastic_net_qite} is re-written in the useful form
\begin{equation}
    \min_{\vec a}\; \frac{1}{2}\vec a^{\,T}A_\rho\vec a
    +\vec b^{\,T}\vec a + \lambda\|\vec a\|_1
\end{equation}
with
\begin{equation}
    A_\rho=A+\rho I,
    \qquad
    A=\boldsymbol{S}+\boldsymbol{S}^{T}=2\Sbar,
\end{equation}
To recover Eq.~\eqref{eq:elastic_net_qite} note that $\frac{1}{2}\vec a^{\,T}A\vec a = \vec a^{\,T}\boldsymbol{S}\vec a$ and the constant factor $\frac{1}{2}$ can be absorbed into $\rho$ to give the ridge term in \eqref{eq:elastic_net_qite}.  The smooth part has gradient
\begin{equation}\label{eq:gradient_smooth}
    A_\rho\vec a+\vec b.
\end{equation}

FISTA applies a gradient step to the smooth part $\frac{1}{2}\vec a^{\,T}A_\rho\vec a +\vec b^{\,T}\vec a$  and the proximal map of the
nondifferentiable term $\lambda\|\vec a\|_1$~\cite{doi:10.1137/080716542}. For the
\(\ell_1\) norm, this proximal map is the componentwise soft-thresholding
operator
\begin{equation}
    \left[\mathcal S_\gamma(\vec z)\right]_I
    =\operatorname{sign}(z_I)\max\{|z_I|-\eta \lambda,0\},
\end{equation}
where $\eta$ is the step size used in FISTA. Starting from \(\vec a_0\), with \(\vec y_0=\vec a_0\) and \(t_0=1\), the
iterations used in the code are
\begin{align}
    \vec a_{r+1}
    &=\mathcal S_{\eta\lambda}\!\left[
        \vec y_r-\eta(A_\rho\vec y_r+\vec b)
      \right],\\
    t_{r+1}
    &=\frac{1+\sqrt{1+4t_r^2}}{2},\\
    \vec y_{r+1}
    &=\vec a_{r+1}
      +\frac{t_r-1}{t_{r+1}}(\vec a_{r+1}-\vec a_r).
\end{align}
Thus each iteration consists of a dense matrix--vector product, a
soft-thresholding step, and the FISTA momentum update. The step size is chosen
as
\begin{equation}
    \eta=\frac{1}{L},
    \qquad
    L=\|A_\rho\|_2,
\end{equation}
because \(L\) is a Lipschitz constant for Eq.~\eqref{eq:gradient_smooth}. The implementation
stops when
\begin{equation}
    \|\vec a_{r+1}-\vec a_r\|_2
    \leq
    \varepsilon_{\mathrm{FISTA}}
    \left(\|\vec a_r\|_2+\delta\right),
\end{equation}
where \(\delta\) is a small numerical safeguard, or when the maximum number
of iterations is reached. When available, the coefficients from the previous
time step for the same Hamiltonian piece are used as a warm start.

The \(\ell_1\) penalty produces exact zeros but also shrinks the surviving
coefficients. To reduce this bias, after convergence the implementation defines the
active set
\begin{equation}
    C=\{I:|a_I|>\varepsilon_{\mathrm{supp}}\}
\end{equation}
and solves the unpenalized quadratic problem restricted to that support,
\begin{equation}
    (A_\rho)_{CC}\vec a_C=\vec q_C,
    \qquad
    a_I=0\quad\text{for }I\notin C.
\end{equation}
This debiasing step keeps the sparse support selected by FISTA while
re-estimating the values of its active coefficients. Consequently, the
debiased vector is the unpenalized solution on the selected support, rather
than the exact minimizer of the original \(\ell_1\)-regularized objective.

\subsection{Why ridge stabilization is needed}
\label{app:ridge_example}

The difficulty created by an ill-conditioned Gram matrix is not merely the presence of exact zero eigenvalues, which could in principle be treated by a Moore-Penrose pseudoinverse. The more important issue is that the QITE coefficient representation is state dependent. Distinct Pauli strings may act identically, or nearly identically, on the current state, so large changes in the coefficient vector can correspond to little or no change in the tangent vector being fitted. Exact null directions lead to nonuniqueness, while near-null directions make the coefficient determination sensitive to finite-shot measurement noise and numerical perturbations.

\paragraph{Illustrative example.}
The need for regularization can be seen from a two-direction example written in terms of the tangent directions of Eq.~\eqref{eqn:tangdir}. Suppose that two distinct Pauli strings \(I,J\in\mathcal P_n\) generate nearly identical tangent directions, so that the Gram-matrix elements \(S_{IJ}=\braket{e_I^{(n)}}{e_J^{(n)}}\) satisfy
\begin{equation}
    \Sbar_{IJ}=\tfrac{1}{2}\bigl(S_{IJ}+S_{JI}\bigr)=r,
    \qquad
    r=1-\varepsilon,
    \qquad
    0<\varepsilon\ll 1.
\end{equation}
Restricted to these two directions, the symmetrized Gram matrix of Eq.~\eqref{eq:real_gram} contains the block
\begin{equation}
    \Sbar\big|_{\{I,J\}}
    =
    \begin{pmatrix}
        1 & r\\
        r & 1
    \end{pmatrix},
\end{equation}
whose eigenvalues are \(1+r\) and \(1-r=\varepsilon\). The near-null direction is proportional to the coefficient difference between the two Pauli strings. Therefore, a small perturbation of the measured data in this direction can be amplified by a factor of order \(1/\varepsilon\) in the fitted coefficients. An exact Moore--Penrose pseudoinverse removes only exactly zero modes; it does not prevent the amplification caused by small but nonzero eigenvalues. A truncated pseudoinverse could discard such near-null modes, but choosing the truncation threshold is itself a form of regularization. Ridge regularization makes this stabilization explicit by replacing the small denominator \(\varepsilon\) with \(\varepsilon+\rho\).

\section{Computational cost analysis}
\label{app:computational-cost}

We summarize the dominant classical post-processing cost of the implementation
used in this work. This analysis begins after the QITE Gram matrix and
right-hand side have been constructed and therefore excludes their quantum
measurement cost.

\subsection{Sparse-QITE}
\label{app:fista_vanilla_cost}

Let \(M\) be the size of the candidate Pauli set for one Hamiltonian piece at
one time step, typically of order $O(\exp(D)$, let \(K_{\mathrm{FISTA}}\) be the number of FISTA iterations,
and let \(s\) be the size of the active set. Each FISTA iteration is dominated
by a dense matrix--vector product and therefore costs \(O(M^2)\). After the
active support has been identified, an optional debiasing step solves the
unpenalized problem on that support, at cost \(O(s^3)\).

The current implementation also computes the operator norm of the dense QITE
matrix to set the FISTA step size. This calculation costs \(O(M^3)\) and is
therefore the dominant operation. The cost of one local Sparse-QITE update is
\begin{equation}
\label{eq:fista_implementation_cost}
    O\!\left(M^3+K_{\mathrm{FISTA}}M^2+s^3\right),
\end{equation}
and peak storage \(O(M^2)\).

As a result, the
leading total cost of the present implementation is therefore
\begin{equation}
    O\!\left(M^3\right).
\end{equation}
This cubic scaling is not intrinsic to FISTA. The explicit norm calculation
can be replaced by a suitable upper bound, a backtracking line search, or an
iterative norm estimate~\cite{doi:10.1137/080716542,golub2013matrix}. In that
case, and when the debiasing solve is restricted to a small active set
\(s\ll M\) or disabled, the dominant Sparse-QITE optimization cost becomes
\(O(M^2)\). Thus Sparse-QITE can be implemented
with quadratic rather than cubic dependence on the full Pauli-pool size.

\subsection{Adaptive Sparse-QITE}
\label{app:fista_adaptive_cost}

Adaptive Sparse-QITE first computes the unregularized QITE update needed to
define the reference value in the residual \(\mathcal R_n(\lambda)\). This
requires solving a dense \(M\times M\) linear system and costs \(O(M^3)\) for
each time step and local Hamiltonian piece. It then evaluates
\(K_\lambda\) candidate values of \(\lambda\). The corresponding grid or
binary search requires a FISTA solve for each candidate and therefore costs
\(O(K_\lambda K_{\mathrm{FISTA}}M^2)\). Once \(\lambda\) has been selected,
the final active solution may be debiased at cost \(O(s^3)\).

The adaptive local-update cost is consequently
\begin{equation}
\label{eq:adaptive_fista_implementation_cost}
    O\!\left(
        M^3+K_\lambda K_{\mathrm{FISTA}}M^2+s^3
    \right),
\end{equation}
and hence the leading cost is
\begin{equation}
    O\!\left(M^3\right).
\end{equation}
Unlike Sparse-QITE, avoiding an explicit operator-norm calculation
does not remove the cubic leading term in the adaptive method, because the
unregularized QITE reference update is still required for the residual.

\section{Adaptive Sparse-QITE}
\label{app:adaptive_binary_search}

We describe the adaptive selection for one local Hamiltonian piece at
imaginary-time step \(n\); the procedure is repeated independently for every
piece, as a result whenever a sub-index $n$ appears it is understood to mean at a local Hamiltonian piece and at certain time step. We first prove the relation between the residual used by the adaptive
rule and the local infidelity, and then establish the upper endpoint and the
properties of the binary search. Suppressing
the update labels, the elastic-net objective in
Eq.~\eqref{eq:elastic_net_qite} can be written as
\begin{equation}
\label{eq:adaptive_elastic_net_objective}
    \mathcal L_{\lambda,\rho}(\vec a)
    =
    f_0
    +\vec b^{\,T}\vec a
    +\vec a^{\,T}\boldsymbol{S}\vec a
    +\rho\|\vec a\|_2^2
    +\lambda\|\vec a\|_1.
\end{equation}

\subsection{Objective residual and local infidelity.}
Let \(\vec a_*\) be a minimizer of the QITE objective \(f\), let
\(\vec a_{\lambda,\rho}\) be the elastic-net solution, and define
\begin{equation}
    \delta\vec a
    :=\vec a_{\lambda,\rho}-\vec a_* ,
    \qquad
    \Delta\hat A
    :=\sum_I\delta a_I\hat\sigma_I.
\end{equation}
The stationarity condition for the unpenalized minimizer is
\begin{equation}
    \vec b+(\boldsymbol{S}+\boldsymbol{S}^T)\vec a_*=0.
\end{equation}
Expanding the quadratic objective about \(\vec a_*\), the linear term
therefore cancels and gives the exact identity
\begin{align}
    \mathcal R(\lambda)
    &:=f(\vec a_{\lambda,\rho})-f(\vec a_*)
    \\
    &=\delta\vec a^{\,T}\boldsymbol{S}\,\delta\vec a
    =\langle\psi|(\Delta\hat A)^2|\psi\rangle.
\label{eq:appendix_residual_gram}
\end{align}
Equation~\eqref{eq:appendix_residual_gram} also shows that the residual
is nonnegative and is independent of the constant term in \(f\).

Both local updates start from the same normalized state \(\ket\psi\). Define
\begin{equation}
    \ket{\phi_*}
    :=e^{-i\Delta\tau\hat A_*}\ket\psi,
    \qquad
    \ket{\phi_{\lambda,\rho}}
    :=e^{-i\Delta\tau\hat A_{\lambda,\rho}}\ket\psi,
\end{equation}
and their local fidelity
\begin{equation}\label{eq:Floc-def}
    F^{\mathrm{loc}}(\lambda)
    :=|\langle\phi_*|\phi_{\lambda,\rho}\rangle|^2.
\end{equation}
Expanding both exponentials through second order gives
\begin{align}
    \langle\phi_*|\phi_{\lambda,\rho}\rangle
    ={}&1-i\Delta\tau\langle\Delta\hat A\rangle
    -\frac{\Delta\tau^2}{2}
      \langle(\Delta\hat A)^2\rangle
    \\
    &+\frac{\Delta\tau^2}{2}
      \langle[\hat A_*,\Delta\hat A]\rangle
    +O(\Delta\tau^3),
\end{align}
where all expectation values are taken in \(\ket\psi\). The commutator
expectation is purely imaginary and hence drops out after taking the modulus
squared. Consequently,
\begin{align}
    1-F^{\mathrm{loc}}(\lambda)
    &=\Delta\tau^2
      \left(
        \langle(\Delta\hat A)^2\rangle
        -\langle\Delta\hat A\rangle^2
      \right)
      +O(\Delta\tau^3)
    \\
    &=\Delta\tau^2
      \left(
        \mathcal R(\lambda)
        -\langle\Delta\hat A\rangle^2
      \right)
      +O(\Delta\tau^3).
\label{eq:appendix_general_residual_infidelity}
\end{align}
Thus, in full generality, local infidelity is proportional to the variance of
the generator difference, rather than to its uncentered second moment. In
particular,
\begin{equation}
    1-F^{\mathrm{loc}}(\lambda)
    \leq
    \Delta\tau^2\mathcal R(\lambda)+O(\Delta\tau^3).
\end{equation}

For the real-valued Hamiltonians and states considered in our numerical
experiments, the QITE pool is restricted to Pauli strings containing an odd
number of \(Y\) operators. Every such Hermitian Pauli string is purely
imaginary in the computational basis, and hence has zero expectation value
in a real state. It follows that
\(\langle\Delta\hat A\rangle=0\), so
Eq.~\eqref{eq:appendix_general_residual_infidelity} reduces to
\begin{equation}
    1-F^{\mathrm{loc}}(\lambda)
    =\Delta\tau^2\mathcal R(\lambda)+O(\Delta\tau^3),
\label{eq:appendix_residual_infidelity_real_pool}
\end{equation}
which proves Eq.~\eqref{eq:adaptive_residual_infidelity}. The tolerance in
Eq.~\eqref{eq:adaptive_lambda_rule} therefore controls the leading local
infidelity directly in this setting. For a general complex state or a general
Pauli pool, Eq.~\eqref{eq:appendix_general_residual_infidelity} is the
appropriate statement. Moreover, Eq.~\eqref{eq:tangent_vector_error_adaptive} follows after we notice that 
\begin{equation}
    \ket{v_{\lambda,\rho}}-\ket v = -i\Delta \hat{A} \ket{\psi},
\end{equation}
where the vectors $\ket{v_{\lambda,\rho}}$ and $\ket v$ are defined in Eq.~\eqref{eq:tangent_vectors}. Taking norms of this gives us the result. 
\subsection{Maximum sparsity beyond $\lambda_{\mathrm{max},n}$}We now establish the search interval. Because the objective
\eqref{eq:adaptive_elastic_net_objective} is convex, \(\vec a=\vec 0\) is
optimal if and only if it satisfies the subgradient optimality condition
\begin{equation}
\label{eq:adaptive_zero_kkt}
    \vec 0
    \in
    \vec b
    +2(\Sbar+\rho I)\vec 0
    +\lambda\,\partial\|\vec 0\|_1
    =
    \vec b+\lambda\,\partial\|\vec 0\|_1.
\end{equation}
The \(\ell_2\) regularizer therefore does not alter the zero-solution
threshold: its gradient, \(2\rho\vec a\), vanishes at the origin. Since
\begin{equation}
    \partial\|\vec 0\|_1
    =
    \left\{
        \vec z:\|\vec z\|_\infty\leq1
    \right\},
\end{equation}
condition~\eqref{eq:adaptive_zero_kkt} holds precisely when there exists a
vector \(\vec z\) with \(\|\vec z\|_\infty\leq1\) such that
\(\vec b+\lambda\vec z=\vec 0\). Equivalently,
\begin{equation}
    |b_I|\leq\lambda
    \quad\text{for every }I,
    \qquad\Longleftrightarrow\qquad
    \lambda\geq\|\vec b\|_\infty.
\end{equation}
The smallest such value is therefore
\begin{equation}
\label{eq:adaptive_lambda_max_proof}
    \lambda_{\max}
    =
    \|\vec b\|_\infty
    =
    \max_{I\in\mathcal P}|b_I|.
\end{equation}
For \(\rho>0\), Eq.~\eqref{eq:adaptive_elastic_net_objective} is strongly
convex, so this zero solution is unique. Thus, the elastic-net term leaves the
LASSO threshold unchanged while removing any possible ambiguity between
multiple minimizers. The desired adaptive regularization strength consequently
lies in \([0,\lambda_{\max}]\).

\subsection{Binary search optimization} We start by computing the QITE endpoint at \(\lambda=0\). If \(\lambda_{\max}\) is also feasible, it is returned immediately,
corresponding to a zero generator that already meets the local tolerance.

In the remaining case, the feasible--infeasible boundary is bracketed by
\(\lambda_{\mathrm{lo}}=0\) and
\(\lambda_{\mathrm{hi}}=\lambda_{\max,n}\). At each iteration, the midpoint
\begin{equation}
    \lambda_{\mathrm{mid}}
    =
    \frac{\lambda_{\mathrm{lo}}+\lambda_{\mathrm{hi}}}{2}
\end{equation}
is tested by solving Eq.~\eqref{eq:elastic_net_qite} and evaluating the
objective \(\mathcal R(\lambda_{\mathrm{mid}})\). If the objective
satisfies the tolerance, the midpoint is feasible and replaces
\(\lambda_{\mathrm{lo}}\); otherwise it replaces
\(\lambda_{\mathrm{hi}}\). Thus, after every update,
\begin{equation}
\label{eq:adaptive_binary_invariant}
    \mathcal R(\lambda_{\mathrm{lo}})
    \leq\varepsilon_{\mathrm{obj}},
    \qquad
    \mathcal R(\lambda_{\mathrm{hi}})
    >\varepsilon_{\mathrm{obj}}.
\end{equation}
The first relation explains why the algorithm returns
\(\lambda_{\mathrm{lo}}\), rather than the midpoint of the final interval:
the returned value is known to satisfy the accuracy constraint. This
bracketing procedure identifies the largest feasible value only when the
feasible set is an interval, i.e., when
\(\mathcal R(\lambda)\leq\varepsilon_{\mathrm{obj}}\) for all
\(0\leq\lambda\leq\widetilde\lambda\). Equivalently, it is sufficient that
\(\mathcal R(\lambda)\) be nondecreasing over the search interval.

\subsection{Monotonicity of \(\mathcal R_n(\lambda)\)}For completeness, the monotonicity implied by convex regularization must be
stated carefully. Define the smooth elastic-net objective
\begin{equation}
    h_\rho(\vec a)
    :=
    f_n(\vec a)+\rho\|\vec a\|_2^2,
\end{equation}
and let \(\vec a_i\) minimize
\(h_\rho(\vec a)+\lambda_i\|\vec a\|_1\), with
\(0\leq\lambda_1<\lambda_2\). Optimality at the two values gives
\begin{align}
    h_\rho(\vec a_1)+\lambda_1\|\vec a_1\|_1
    &\leq
    h_\rho(\vec a_2)+\lambda_1\|\vec a_2\|_1,\\
    h_\rho(\vec a_2)+\lambda_2\|\vec a_2\|_1
    &\leq
    h_\rho(\vec a_1)+\lambda_2\|\vec a_1\|_1.
\end{align}
Adding these inequalities shows that
\(\|\vec a_2\|_1\leq\|\vec a_1\|_1\); substituting this result into the first
inequality yields
\begin{equation}
\label{eq:adaptive_smooth_objective_monotonicity}
    h_\rho(\vec a_1)
    \leq
    h_\rho(\vec a_2).
\end{equation}
Therefore, convexity guarantees that
\(f_n(\vec a_{\lambda,\rho})+\rho\|\vec a_{\lambda,\rho}\|_2^2\) is
nondecreasing along the elastic-net path. For \(\rho=0\), this proves the
required monotonicity of the objective
\(\mathcal R(\lambda)\). For \(\rho>0\), however,
Eq.~\eqref{eq:adaptive_smooth_objective_monotonicity} does not by itself prove
that \(f(\vec a_{\lambda,\rho})\) is nondecreasing after the ridge term is
removed.
Binary search based on $\mathcal R(\lambda)$ therefore requires the additional assumption that the unpenalized QITE mismatch remains nondecreasing along the elastic-net regularization path. This property is guaranteed in the strict LASSO limit $\rho=0$, but does not follow from convexity alone when $\rho>0$. Nevertheless, since the ridge parameter used here, $\rho=10^{-10}$, serves only as a small numerical stabilization, we retain binary search as an efficient procedure for locating the largest feasible value of $\lambda$. 

\subsection{Numerical details}The implementation performs binary search over a finite logarithmic grid,
rather than over the continuous interval \([0,\lambda_{\max}]\). For each
local update, we construct the ordered candidate set
\begin{equation}
\label{eq:adaptive_lambda_grid}
    \Lambda
    =
    \left\{
        \lambda_{\max}
        10^{-8j/(N_\lambda-1)}
    \right\}_{j=0}^{N_\lambda-1}
    \cup\{0\},
\end{equation}
where the positive candidates decrease from \(\lambda_{\max}\) to
\(10^{-8}\lambda_{\max}\). Assuming feasibility is monotone along this
ordered grid, the largest feasible candidate is the first feasible grid entry.
The binary search maintains an infeasible lower index and a feasible upper
index and repeatedly tests the integer midpoint between them. It terminates
when the two indices are adjacent, at which point the feasible index identifies
the largest grid value known to satisfy the tolerance.

If the grid contains \(N_\lambda\) positive candidates and the appended zero,
the initial index separation is \(N_\lambda\). Each midpoint evaluation
reduces the current separation \(d\) to at most \(\lceil d/2\rceil\), so at
most
\begin{equation}
\label{eq:adaptive_binary_iteration_count}
    K_{\mathrm{bin}}
    =
    \left\lceil\log_2 N_\lambda\right\rceil
\end{equation}
midpoint evaluations are required. In the present implementation,
\(N_\lambda=25\), so the search requires at most five midpoint evaluations,
in addition to the endpoint checks. Each midpoint evaluation entails one
elastic-net solve and one evaluation of \(\mathcal R\). The closest
previously evaluated grid point supplies the initial iterate for FISTA. This
warm start does not alter the target optimization problem; provided the same
convergence criterion is reached, the strongly convex elastic-net objective
has the same unique minimizer. No quantitative runtime improvement from warm
starting is assumed here. The Hamiltonian-piece index has been omitted
throughout only to simplify the notation.

\section{Local regularization bounds}
\label{app:local_regularization_bounds}

\paragraph{Proof of Proposition~\ref{prop:local_same_set_bound}.}
The argument is local to a fixed update $(n,k)$. We therefore suppress the
labels $(n,k)$ and write $\mathcal P$ for the fixed candidate Pauli set,
$M=|\mathcal P|$, and $\Sbar$ for the corresponding symmetrized Gram matrix of Eq.~\eqref{eq:real_gram}. Let
\begin{equation}
    \kappa:=\lambda_{\min}(\Sbar)\ge0,
    \qquad
    \Sbar\succeq\kappa \boldsymbol{\hat{I}}.
\end{equation}
Let $\vec b$ be the linear coefficient vector in the unscaled quadratic
QITE objective and define
\begin{equation}
    \vec d:=-\frac{1}{2}\vec b.
\end{equation}
The ridge-stabilized coefficients satisfy
\begin{equation}
\label{eq:appendix_ridge_normal_equation}
    (\Sbar+\rho \boldsymbol{\hat{I}})\vec a_{\rho}=\vec d.
\end{equation}
For
$\vec z_{\lambda,\rho}\in
\partial\|\vec a_{\lambda,\rho}\|_1$, define
\begin{equation}
    \vec d_{\lambda,\rho}
    :=
    \vec d-\frac{\lambda}{2}\vec z_{\lambda,\rho}.
\end{equation}
The elastic-net optimality condition is
\begin{equation}
\label{eq:appendix_elastic_net_kkt}
    (\Sbar+\rho \boldsymbol{\hat{I}})\vec a_{\lambda,\rho}
    =
    \vec d_{\lambda,\rho}.
\end{equation}
Subtracting Eq.~\eqref{eq:appendix_ridge_normal_equation} from
Eq.~\eqref{eq:appendix_elastic_net_kkt} gives
\begin{equation}
\label{eq:appendix_same_set_subtraction}
    (\Sbar+\rho \boldsymbol{\hat{I}})
    (\vec a_{\lambda,\rho}-\vec a_{\rho})
    =
    -\frac{\lambda}{2}\vec z_{\lambda,\rho}.
\end{equation}
Since $\Sbar\succeq\kappa I$ and $\rho>0$,
\begin{equation}
    \Sbar+\rho \boldsymbol{\hat{I}}\succeq(\kappa+\rho)\boldsymbol{\hat{I}},
\end{equation}
and therefore
\begin{equation}
\label{eq:appendix_ridge_inverse_bound}
    \|(\Sbar+\rho \boldsymbol{\hat{I}})^{-1}\|_{\rm op}
    \le
    \frac{1}{\kappa+\rho}.
\end{equation}
Each component of $\vec z_{\lambda,\rho}$ has magnitude at most one, so
\begin{equation}
    \|\vec z_{\lambda,\rho}\|_2^2
    =
    \sum_{I\in\mathcal P}|z_I|^2
    \le M.
\end{equation}
Taking norms in Eq.~\eqref{eq:appendix_same_set_subtraction} yields
\begin{equation}
    \|\vec a_{\lambda,\rho}-\vec a_{\rho}\|_2
    \le
    \frac{\lambda\sqrt M}{2(\kappa+\rho)},
\end{equation}
which proves Eq.~\eqref{eq:same_set_coeff_bound}.

Now, set
\begin{equation}
    \delta\vec a
    :=
    \vec a_{\lambda,\rho}-\vec a_{\rho}.
\end{equation}
The squared tangent-vector difference is the corresponding Gram norm,
\begin{equation}
    \|\ket{v_{\lambda,\rho}}-\ket{v_{\rho}}\|^2
    =
    \delta\vec a^{\,T}\Sbar\,\delta\vec a.
\end{equation}
Since $\Sbar\preceq \Sbar+\rho \boldsymbol{\hat{I}}$, Eq.~\eqref{eq:appendix_same_set_subtraction}
implies
\begin{align}
    \|\ket{v_{\lambda,\rho}}-\ket{v_{\rho}}\|^2
    &\le
    \delta\vec a^{\,T}(\Sbar+\rho \boldsymbol{\hat{I}})\delta\vec a
    \\
    &=
    \frac{\lambda^2}{4}
    \vec z_{\lambda,\rho}^{\,T}
    (\Sbar+\rho \boldsymbol{\hat{I}})^{-1}
    \vec z_{\lambda,\rho}
    \\
    &\le
    \frac{\lambda^2M}{4(\kappa+\rho)}.
\end{align}
Taking the square root proves Eq.~\eqref{eq:same_set_tangent_bound}.

\paragraph{Proof of Corollary~\ref{cor:ordinary_reference_bound}.}
The ordinary-QITE coefficients minimize the real quadratic least-squares
objective. Its first-order optimality condition is
\begin{equation}
    \Sbar\vec a=\vec d.
\end{equation}
Because this is the normal equation of the ordinary-QITE least-squares
problem, no separate consistency assumption is imposed. We choose the
Moore--Penrose solution
\begin{equation}
    \vec a=\Sbar^\dagger\vec d,
    \qquad
    B:=\|\vec a\|_2,
\end{equation}
as the minimum-norm representative. If \(\widetilde{\vec a}\) is another
ordinary-QITE minimizer, then
\(\vec u:=\widetilde{\vec a}-\vec a\in\ker \Sbar\), and hence
\begin{equation}
    \left\|
    \sum_{I\in\mathcal P}u_I\ket{e_I}
    \right\|^2
    =
    \vec u^{\,T}\Sbar\vec u
    =0.
\end{equation}
Thus the ordinary-QITE tangent vector is independent of the chosen
least-squares representative, even when the coefficient vector is
nonunique. Subtracting $\Sbar\vec a=\vec d$ from
Eq.~\eqref{eq:appendix_elastic_net_kkt} gives
\begin{equation}
\label{eq:appendix_ordinary_subtraction}
    (\Sbar+\rho I)(\vec a_{\lambda,\rho}-\vec a)
    =
    -\frac{\lambda}{2}\vec z_{\lambda,\rho}
    -
    \rho\vec a.
\end{equation}
With
$\delta\vec a:=\vec a_{\lambda,\rho}-\vec a$,
Eq.~\eqref{eq:appendix_ridge_inverse_bound},
$\|\vec z_{\lambda,\rho}\|_2\le\sqrt M$, and
$\|\vec a\|_2=B$ imply
\begin{equation}
    \|\delta\vec a\|_2
    \le
    \frac{\frac{\lambda}{2}\sqrt M+\rho B}{\kappa+\rho},
\end{equation}
which proves Eq.~\eqref{eq:same_set_ordinary_coeff_bound}. Moreover,
\begin{equation}
    \|\ket{v_{\lambda,\rho}}-\ket v\|^2
    =
    \delta\vec a^{\,T}\Sbar\,\delta\vec a,
\end{equation}
and Eq.~\eqref{eq:appendix_ordinary_subtraction} gives
\begin{align}
    \|\ket{v_{\lambda,\rho}}-\ket v\|^2
    &\le
    \delta\vec a^{\,T}(\Sbar+\rho I)\delta\vec a
    \\
    &=
    -\frac{\lambda}{2}\delta\vec a^{\,T}\vec z_{\lambda,\rho}
    -
    \rho\delta\vec a^{\,T}\vec a
    \\
    &\le
    \left(\frac{\lambda}{2}\sqrt M+\rho B\right)
    \|\delta\vec a\|_2
    \\
    &\le
    \frac{\left(\frac{\lambda}{2}\sqrt M+\rho B\right)^2}
    {\kappa+\rho}.
\end{align}
Taking the square root proves
Eq.~\eqref{eq:same_set_ordinary_reference_bound}.

\section{Stability and trajectory-level error propagation}
\label{app:trajectory_stability}

We first state the stability property of the ridge-QITE update used in the
trajectory analysis of Sec.~\ref{subsec:error-bounds}. Let $F_{\rho}$ denote
the exact ridge-QITE one-step map for a fixed local update.

\begin{lemma}[Stability of the ridge-QITE update]
\label{lem:ridge_qite_stability}
For fixed finite $\mathcal P$ and $\rho>0$, there exists a finite constant
$L_{\rho}$ such that
\begin{equation}
\label{eq:ridge_qite_stability}
    \|F_{\rho}(\phi)-F_{\rho}(\varphi)\|
    \le
    \left(1+L_{\rho}\Delta\tau\right)
    \|\ket\phi-\ket\varphi\|.
\end{equation}
\end{lemma}

\paragraph{Proof of Lemma~\ref{lem:ridge_qite_stability}.}
Fix a local update $(n,k)$, its candidate Pauli set $\mathcal P_{n,k}$, and
$\rho>0$. As in the main-text local analysis, we suppress both fixed update
labels and write
\begin{equation}
    \mathcal P:=\mathcal P_{n,k},
    \qquad
    M:=|\mathcal P|.
\end{equation}
For a normalized input state $\ket\phi$, define
\begin{equation}
    \Sbar(\phi):=\Sbar_{\mathcal P_{n,k}}^{(n,k)}(\phi).
\end{equation}
Because the candidate Pauli set $\mathcal P$ is finite, the entries of
$\Sbar(\phi)$ and the components of $\vec d(\phi)$ are finite algebraic
combinations of expectation values of bounded operators. In particular,
for normalized states and every bounded operator $\hat O$,
\begin{equation}
\label{eq:expectation_lipschitz}
    \left|
    \bra\phi\hat O\ket\phi
    -
    \bra\varphi\hat O\ket\varphi
    \right|
    \le
    2\|\hat O\|_{\rm op}\|\ket\phi-\ket\varphi\|.
\end{equation}
The normalization factor entering $\vec d(\phi)$ is also bounded away
from zero for the fixed bounded Hamiltonian piece $\boldsymbol{\hat h}_k$, since
\begin{equation}
    c(\phi)
    :=
    \sqrt{\bra\phi e^{-2\Delta\tau\boldsymbol{\hat h}_k}\ket\phi}
    \ge
    e^{-\Delta\tau\|\boldsymbol{\hat h}_k\|_{\rm op}}
    >0.
\end{equation}
It follows directly from these estimates, and from the finiteness of
$\mathcal P$, that there exist finite constants $B_d$, $L_d$, and $L_S$
such that
\begin{align}
    \|\vec d(\phi)\|_2
    &\le B_d,
    \label{eq:uniform_d_bound}
    \\
    \|\vec d(\phi)-\vec d(\varphi)\|_2
    &\le
    L_d\|\ket\phi-\ket\varphi\|,
    \label{eq:d_lipschitz_bound}
    \\
    \|\Sbar(\phi)-\Sbar(\varphi)\|_{\rm op}
    &\le
    L_S\|\ket\phi-\ket\varphi\|.
    \label{eq:S_lipschitz_bound}
\end{align}

Explicit admissible choices for these constants, including versions uniform
along a finite ordered trajectory, are derived separately in
App.~\ref{app:explicit_uniform_constants}.

Let
\begin{equation}
    G_{\rho}(\phi):=\Sbar(\phi)+\rho I,
    \qquad
    \vec a_{\rho}(\phi)
    :=
    G_{\rho}(\phi)^{-1}\vec d(\phi).
\end{equation}
Since $\Sbar(\phi)\succeq0$,
\begin{equation}
    G_{\rho}(\phi)\succeq\rho I,
    \qquad
    \|G_{\rho}(\phi)^{-1}\|_{\rm op}\le\frac1\rho.
\end{equation}
For two normalized states, add and subtract
$G_{\rho}(\phi)^{-1}\vec d(\varphi)$ to obtain
\begin{align}
    &\vec a_{\rho}(\phi)-\vec a_{\rho}(\varphi)
    \\
    ={}&
    G_{\rho}(\phi)^{-1}
    [\vec d(\phi)-\vec d(\varphi)]
    \\
    &+
    G_{\rho}(\phi)^{-1}
    [G_{\rho}(\varphi)-G_{\rho}(\phi)]
    \vec a_{\rho}(\varphi).
\end{align}
Moreover,
\begin{equation}
    \|\vec a_{\rho}(\varphi)\|_2\le\frac{B_d}{\rho}.
\end{equation}
Therefore
\begin{equation}
\label{eq:ridge_coeff_state_lipschitz}
    \|\vec a_{\rho}(\phi)-\vec a_{\rho}(\varphi)\|_2
    \le
    \left(
    \frac{L_d}{\rho}
    +
    \frac{L_SB_d}{\rho^2}
    \right)
    \|\ket\phi-\ket\varphi\|.
\end{equation}
The corresponding ridge-QITE generator is
\begin{equation}
    \boldsymbol{\hat{A}}_{\rho}(\phi)
    =
    \sum_{I\in\mathcal P}
    a_{\rho,I}(\phi)\boldsymbol{\hat{\sigma}}_I.
\end{equation}
Since each Pauli string has operator norm one,
\begin{align}
    &\|\boldsymbol{\hat{A}}_{\rho}(\phi)
    -\boldsymbol{\hat{A}}_{\rho}(\varphi)\|_{\rm op}
    \\
    &\qquad\le
    \sqrt M\,
    \|\vec a_{\rho}(\phi)-\vec a_{\rho}(\varphi)\|_2.
\end{align}
Consequently,
\begin{equation}
\label{eq:ridge_generator_lipschitz}
    \|\boldsymbol{\hat{A}}_{\rho}(\phi)
    -\boldsymbol{\hat{A}}_{\rho}(\varphi)\|_{\rm op}
    \le
    L_{\rho}\|\ket\phi-\ket\varphi\|,
\end{equation}
where one may take
\begin{equation}
\label{eq:L_rho_definition}
    L_{\rho}
    :=
    \sqrt M
    \left(
    \frac{L_d}{\rho}
    +
    \frac{L_SB_d}{\rho^2}
    \right).
\end{equation}
Define
\begin{equation}
    F_{\rho}(\phi)
    :=
    e^{-i\Delta\tau\boldsymbol{\hat{A}}_{\rho}(\phi)}\ket\phi.
\end{equation}
For completeness, write
\begin{equation*}
    U_{\phi}:=
    e^{-i\Delta\tau\boldsymbol{\hat{A}}_{\rho}(\phi)},
    \qquad
    U_{\varphi}:=
    e^{-i\Delta\tau\boldsymbol{\hat{A}}_{\rho}(\varphi)}.
\end{equation*}
Adding and subtracting $U_{\phi}\ket\varphi$ yields
\begin{align}
    \|F_{\rho}(\phi)-F_{\rho}(\varphi)\|
    &=
    \|U_{\phi}\ket\phi-U_{\varphi}\ket\varphi\|
    \\
    &\le
    \|U_{\phi}(\ket\phi-\ket\varphi)\|
    +
    \|(U_{\phi}-U_{\varphi})\ket\varphi\|.
\end{align}
The first term is exactly
\begin{equation*}
    \|U_{\phi}(\ket\phi-\ket\varphi)\|
    =
    \|\ket\phi-\ket\varphi\|,
\end{equation*}
because $U_{\phi}$ is unitary. For the second term, normalizedness of
$\ket\varphi$ gives
\begin{equation}
\label{eq:stability_second_term_operator_norm}
    \|(U_{\phi}-U_{\varphi})\ket\varphi\|
    \le
    \|U_{\phi}-U_{\varphi}\|_{\rm op}.
\end{equation}
For Hermitian $A$ and $B$, Duhamel's identity reads
\begin{equation}
\label{eq:duhamel_identity_unitaries}
    e^{-itA}-e^{-itB}
    =
    -i\int_0^t
    e^{-i(t-s)A}(A-B)e^{-isB}\,ds.
\end{equation}
Since both exponential factors in the integrand are unitary,
\begin{align}
\label{eq:duhamel_unitary_bound}
    \|e^{-itA}-e^{-itB}\|_{\rm op}
    &\le
    \int_0^t
    \|e^{-i(t-s)A}\|_{\rm op}
    \|A-B\|_{\rm op}
    \|e^{-isB}\|_{\rm op}\,ds
    \\
    &=
    t\|A-B\|_{\rm op}.
\end{align}
Applying this with
$A=\boldsymbol{\hat{A}}_{\rho}(\phi)$,
$B=\boldsymbol{\hat{A}}_{\rho}(\varphi)$, and
$t=\Delta\tau$, and then using
Eq.~\eqref{eq:ridge_generator_lipschitz}, gives
\begin{align}
    \|U_{\phi}-U_{\varphi}\|_{\rm op}
    &\le
    \Delta\tau
    \|\boldsymbol{\hat{A}}_{\rho}(\phi)
    -\boldsymbol{\hat{A}}_{\rho}(\varphi)\|_{\rm op}
    \\
    &\le
    L_{\rho}\Delta\tau
    \|\ket\phi-\ket\varphi\|.
\end{align}
Combining these estimates therefore gives
\begin{align}
    \|F_{\rho}(\phi)-F_{\rho}(\varphi)\|
    &\le
    \|\ket\phi-\ket\varphi\|
    +
    L_{\rho}\Delta\tau
    \|\ket\phi-\ket\varphi\|
    \\
    &=
    \left(1+L_{\rho}\Delta\tau\right)
    \|\ket\phi-\ket\varphi\|.
\end{align}
This proves Lemma~\ref{lem:ridge_qite_stability}.

\paragraph{Proof of Corollary~\ref{cor:trajectory_bound}.}
As in the main text, $r=1,\ldots,R$ labels the ordered local updates. For
each update, define
\begin{align*}
    F^{(r)}(\phi)
    &:=
    e^{-i\Delta\tau\boldsymbol{\hat{A}}^{(r)}(\phi)}\ket\phi,
    \\
    F_{\rho}^{(r)}(\phi)
    &:=
    e^{-i\Delta\tau\boldsymbol{\hat{A}}_{\rho}^{(r)}(\phi)}\ket\phi,
    \\
    F_{\lambda,\rho}^{(r)}(\phi)
    &:=
    e^{-i\Delta\tau
    \boldsymbol{\hat{A}}_{\lambda,\rho}^{(r)}(\phi)}\ket\phi,
\end{align*}
where
$\boldsymbol{\hat{A}}^{(r)}:=
\boldsymbol{\hat{A}}_{n_r,k_r}$ is the ordinary-QITE generator constructed
from the minimum-norm representative of the ordinary-QITE least-squares
solution, in the notation established before the bounds. Whenever two maps
are compared at a common input state, their generators are constructed on
the same set $\mathcal P_{r}$.

We next spell out how the local tangent-vector estimates give the
same-input one-step bounds. For any Hermitian generator $X$ and normalized
state $\ket\phi$, Taylor's formula with integral remainder gives
\begin{align}
\label{eq:unitary_first_order_integral_remainder}
    e^{-i\Delta\tau X}\ket\phi
    &=
    \ket\phi-i\Delta\tau X\ket\phi
    +\ket{R_X(\phi)},
    \\
    \ket{R_X(\phi)}
    &:=-
    \int_0^{\Delta\tau}
    (\Delta\tau-s)X^2e^{-isX}\ket\phi\,ds.
\end{align}
Because $X$ is Hermitian, $e^{-isX}$ is unitary, and hence
\begin{equation}
\label{eq:unitary_first_order_remainder_bound}
    \|\ket{R_X(\phi)}\|
    \le
    \int_0^{\Delta\tau}
    (\Delta\tau-s)\|X\|_{\rm op}^2\,ds
    =
    \frac{(\Delta\tau)^2}{2}\|X\|_{\rm op}^2.
\end{equation}
For the Sparse-QITE--to--ridge-QITE comparison, applying
Eq.~\eqref{eq:unitary_first_order_integral_remainder} to the two generators
at the same input state yields
\begin{align}
\label{eq:sparse_ridge_first_order_decomposition}
    &F_{\lambda,\rho}^{(r)}(\phi)
    -F_{\rho}^{(r)}(\phi)
    \\
    &\quad=
    -i\Delta\tau
    \left(
    \boldsymbol{\hat{A}}_{\lambda,\rho}^{(r)}(\phi)
    -
    \boldsymbol{\hat{A}}_{\rho}^{(r)}(\phi)
    \right)\ket\phi
    +
    \ket{R_{\lambda,\rho}^{(r)}(\phi)}
    -
    \ket{R_{\rho}^{(r)}(\phi)}.
\end{align}
By the definition $\ket{e_I}=-i\boldsymbol{\hat{\sigma}}_I\ket\phi$,
\begin{equation}
\label{eq:generator_difference_equals_tangent_difference}
    -i
    \left(
    \boldsymbol{\hat{A}}_{\lambda,\rho}^{(r)}(\phi)
    -
    \boldsymbol{\hat{A}}_{\rho}^{(r)}(\phi)
    \right)\ket\phi
    =
    \ket{v_{\lambda,\rho}^{(r)}(\phi)}
    -
    \ket{v_{\rho}^{(r)}(\phi)}.
\end{equation}
Therefore Proposition~\ref{prop:local_same_set_bound}, specifically
Eq.~\eqref{eq:same_set_tangent_bound} with the update labels restored, gives
\begin{equation}
\label{eq:sparse_ridge_tangent_q_bound}
    \left\|
    -i
    \left(
    \boldsymbol{\hat{A}}_{\lambda,\rho}^{(r)}(\phi)
    -
    \boldsymbol{\hat{A}}_{\rho}^{(r)}(\phi)
    \right)\ket\phi
    \right\|
    \le
    q_{\lambda,\rho,\mathrm{sp}}^{(r)}(\phi).
\end{equation}
Combining Eqs.~\eqref{eq:unitary_first_order_remainder_bound},
\eqref{eq:sparse_ridge_first_order_decomposition}, and
\eqref{eq:sparse_ridge_tangent_q_bound} gives
\begin{align}
\label{eq:local_sparse_ridge_unitary}
    \|F_{\lambda,\rho}^{(r)}(\phi)
    -F_{\rho}^{(r)}(\phi)\|
    &\le
    \Delta\tau
    q_{\lambda,\rho,\mathrm{sp}}^{(r)}(\phi)
    \\
    &\quad+
    \frac{(\Delta\tau)^2}{2}
    \left(
    \|\boldsymbol{\hat{A}}_{\lambda,\rho}^{(r)}(\phi)\|_{\rm op}^2
    +
    \|\boldsymbol{\hat{A}}_{\rho}^{(r)}(\phi)\|_{\rm op}^2
    \right)
    \\
    &\le
    \Delta\tau
    q_{\lambda,\rho,\mathrm{sp}}^{(r)}(\phi)
    +
    \Gamma_{\mathrm{sp}}(\Delta\tau)^2.
\end{align}
Here one may take
\begin{equation}
\label{eq:Gamma_sp_definition}
    \Gamma_{\mathrm{sp}}
    :=
    \frac12
    \sup_{r,\phi}
    \left(
    \|\boldsymbol{\hat{A}}_{\lambda,\rho}^{(r)}(\phi)\|_{\rm op}^2
    +
    \|\boldsymbol{\hat{A}}_{\rho}^{(r)}(\phi)\|_{\rm op}^2
    \right),
\end{equation}
where the supremum is over the ordered updates and normalized input states
under consideration.

The ridge-QITE--to--ordinary-QITE comparison is identical. In this case,
Corollary~\ref{cor:ordinary_reference_bound} with $\lambda=0$, specifically
Eq.~\eqref{eq:same_set_ordinary_reference_bound}, gives
\begin{equation}
\label{eq:ridge_ordinary_tangent_q_bound}
    \left\|
    -i
    \left(
    \boldsymbol{\hat{A}}_{\rho}^{(r)}(\phi)
    -
    \boldsymbol{\hat{A}}^{(r)}(\phi)
    \right)\ket\phi
    \right\|
    \le
    q_{\rho,\mathrm{ridge}}^{(r)}(\phi).
\end{equation}
Repeating the preceding first-order expansion therefore yields
\begin{align}
\label{eq:local_ridge_ordinary_unitary}
    \|F_{\rho}^{(r)}(\phi)
    -F^{(r)}(\phi)\|
    &\le
    \Delta\tau
    q_{\rho,\mathrm{ridge}}^{(r)}(\phi)
    \\
    &\quad+
    \frac{(\Delta\tau)^2}{2}
    \left(
    \|\boldsymbol{\hat{A}}_{\rho}^{(r)}(\phi)\|_{\rm op}^2
    +
    \|\boldsymbol{\hat{A}}^{(r)}(\phi)\|_{\rm op}^2
    \right)
    \\
    &\le
    \Delta\tau
    q_{\rho,\mathrm{ridge}}^{(r)}(\phi)
    +
    \Gamma_{\mathrm{ridge}}(\Delta\tau)^2,
\end{align}
with the admissible uniform choice
\begin{equation}
\label{eq:Gamma_ridge_definition}
    \Gamma_{\mathrm{ridge}}
    :=
    \frac12
    \sup_{r,\phi}
    \left(
    \|\boldsymbol{\hat{A}}_{\rho}^{(r)}(\phi)\|_{\rm op}^2
    +
    \|\boldsymbol{\hat{A}}^{(r)}(\phi)\|_{\rm op}^2
    \right).
\end{equation}
Thus the combined local first-order remainder appearing in the statement of
the corollary is bounded by
\begin{equation}
    \Gamma(\Delta\tau)^2,
    \qquad
    \Gamma:=\Gamma_{\mathrm{sp}}+\Gamma_{\mathrm{ridge}}.
\end{equation}

Define the two trajectories recursively by
\begin{equation}
    \ket{\Phi_r^{\boldsymbol{\mathcal P}}}
    =
    F^{(r)}
    \!\left(\Phi_{r-1}^{\boldsymbol{\mathcal P}}\right),
    \qquad
    \ket{\Phi_r^{\lambda,\rho,\boldsymbol{\mathcal P}}}
    =
    F_{\lambda,\rho}^{(r)}
    \!\left(\Phi_{r-1}^{\lambda,\rho,\boldsymbol{\mathcal P}}\right),
\end{equation}
with the same initial state, and let
\begin{equation}
    \delta_r
    :=
    \left\|
    \ket{\Phi_r^{\lambda,\rho,\boldsymbol{\mathcal P}}}
    -
    \ket{\Phi_r^{\boldsymbol{\mathcal P}}}
    \right\|.
\end{equation}
Inserting the ridge map twice gives
\begin{align}
    \delta_r
    \le{}&
    \left\|
    F_{\lambda,\rho}^{(r)}
    \!\left(\Phi_{r-1}^{\lambda,\rho,\boldsymbol{\mathcal P}}\right)
    -
    F_{\rho}^{(r)}
    \!\left(\Phi_{r-1}^{\lambda,\rho,\boldsymbol{\mathcal P}}\right)
    \right\|
    \\
    &+
    \left\|
    F_{\rho}^{(r)}
    \!\left(\Phi_{r-1}^{\lambda,\rho,\boldsymbol{\mathcal P}}\right)
    -
    F_{\rho}^{(r)}
    \!\left(\Phi_{r-1}^{\boldsymbol{\mathcal P}}\right)
    \right\|
    \\
    &+
    \left\|
    F_{\rho}^{(r)}
    \!\left(\Phi_{r-1}^{\boldsymbol{\mathcal P}}\right)
    -
    F^{(r)}
    \!\left(\Phi_{r-1}^{\boldsymbol{\mathcal P}}\right)
    \right\|.
\end{align}
The first and third terms are controlled by
Eqs.~\eqref{eq:local_sparse_ridge_unitary} and
\eqref{eq:local_ridge_ordinary_unitary}; the middle term is controlled by
Lemma~\ref{lem:ridge_qite_stability}, with the full update label restored.
Hence
\begin{equation}
\label{eq:appendix_trajectory_recursion}
    \delta_r
    \le
    \left(1+L_{\rho}^{(r)}\Delta\tau\right)\delta_{r-1}
    +
    \Delta\tau\,\widetilde q_{\lambda,\rho}^{(r)}
    +
    \Gamma(\Delta\tau)^2,
\end{equation}
where
\begin{equation}
    \widetilde q_{\lambda,\rho}^{(r)}
    :=
    q_{\lambda,\rho,\mathrm{sp}}^{(r)}
    \!\left(\Phi_{r-1}^{\lambda,\rho,\boldsymbol{\mathcal P}}\right)
    +
    q_{\rho,\mathrm{ridge}}^{(r)}
    \!\left(\Phi_{r-1}^{\boldsymbol{\mathcal P}}\right).
\end{equation}
Thus $\delta_{r-1}$ is the discrepancy accumulated at earlier updates,
whereas the last two terms are the new local contribution at update
$r$. Starting from $\delta_0=0$, iteration gives
\begin{equation}
\label{eq:appendix_iterated_trajectory_bound}
    \delta_R
    \le
    \sum_{r=1}^{R}
    \left[
    \prod_{j=r+1}^{R}
    \left(1+L_{\rho}^{(j)}\Delta\tau\right)
    \right]
    \left(
    \Delta\tau\,\widetilde q_{\lambda,\rho}^{(r)}
    +
    \Gamma(\Delta\tau)^2
    \right).
\end{equation}
Using the elementary inequality
\begin{equation}
    1+x\le e^x,
\end{equation}
with $x=L_{\rho}^{(j)}\Delta\tau\ge0$, we have
\begin{equation}
\label{eq:appendix_product_exponential_bound}
    \prod_{j=r+1}^{R}
    \left(1+L_{\rho}^{(j)}\Delta\tau\right)
    \le
    \exp\!\left[
    \Delta\tau
    \sum_{j=r+1}^{R}L_{\rho}^{(j)}
    \right].
\end{equation}
Consequently, with
\begin{equation}
    \Lambda_R
    :=
    \Delta\tau
    \sum_{j=1}^{R}L_{\rho}^{(j)},
\end{equation}
we obtain
\begin{equation}
\label{eq:appendix_exponential_trajectory_bound}
    \delta_R
    \le
    e^{\Lambda_R}
    \left[
    \Delta\tau
    \sum_{r=1}^{R}
    \widetilde q_{\lambda,\rho}^{(r)}
    +
    \Gamma R(\Delta\tau)^2
    \right].
\end{equation}
Equivalently,
\begin{equation}
    \delta_R
    \le
    e^{\Lambda_R}
    \Delta\tau
    \sum_{r=1}^{R}
    \widetilde q_{\lambda,\rho}^{(r)}
    +
    \mathcal O\!\left(e^{\Lambda_R}R(\Delta\tau)^2\right),
\end{equation}
which is Eq.~\eqref{eq:qite_sparse_trajectory_bound}. Together with
Eq.~\eqref{eq:appendix_iterated_trajectory_bound}, this proves
Corollary~\ref{cor:trajectory_bound}. The corresponding comparison with
imaginary-time evolution follows separately by combining this trajectory
estimate with the ITE-to-QITE bounds of Ref.~\cite{Motta2019} through the
triangle inequality.

\section{Explicit admissible constants for the stability bounds}
\label{app:explicit_uniform_constants}

This appendix records explicit state-independent choices for the constants in
Eqs.~\eqref{eq:uniform_d_bound}--\eqref{eq:S_lipschitz_bound}. The purpose is
only to justify the uniform estimates used in the stability proof; the
constants below are convenient rather than optimized.

The following choices are uniform over all normalized input states for this
fixed local update; they are convenient rather than optimized. Set
\begin{equation*}
    \tau:=\Delta\tau,
    \qquad
    H:=\|\boldsymbol{\hat h}_k\|_{\rm op},
    \qquad
    E:=e^{-\tau\boldsymbol{\hat h}_k},
\end{equation*}
and define
\begin{equation*}
    C_I:=\|[\boldsymbol{\hat h}_k,\boldsymbol{\hat{\sigma}}_I]\|_{\rm op},
    \qquad
    C_{\mathcal P}:=
    \left(\sum_{I\in\mathcal P}C_I^2\right)^{1/2}
    \le 2\sqrt M\,H.
\end{equation*}
For the symmetrized Gram matrix of Eq.~\eqref{eq:real_gram},
\begin{equation*}
    \Sbar_{IJ}(\phi)
    =
    \bra\phi\boldsymbol{\hat O}_{IJ}\ket\phi,
    \qquad
    \boldsymbol{\hat O}_{IJ}
    :=
    \frac12\left(
    \boldsymbol{\hat{\sigma}}_I\boldsymbol{\hat{\sigma}}_J
    +
    \boldsymbol{\hat{\sigma}}_J\boldsymbol{\hat{\sigma}}_I
    \right),
\end{equation*}
with \(\|\boldsymbol{\hat O}_{IJ}\|_{\rm op}\le1\). Hence, writing
\(\delta:=\|\ket\phi-\ket\varphi\|\),
\begin{align*}
    \|\Sbar(\phi)-\Sbar(\varphi)\|_{\rm op}
    &\le
    \|\Sbar(\phi)-\Sbar(\varphi)\|_{F}
    \\
    &\le
    \left(\sum_{I,J}(2\delta)^2\right)^{1/2}
    =2M\delta,
\end{align*}
so one may take \(L_S=2M\).

For \(\vec d\), define the bounded Hermitian operator
\begin{equation*}
    \boldsymbol{\hat K}_I
    :=
    -i[E,\boldsymbol{\hat{\sigma}}_I],
    \qquad
    d_I(\phi)
    =
    \frac{\bra\phi\boldsymbol{\hat K}_I\ket\phi}
    {2\tau c(\phi)}.
\end{equation*}
Thus \(\Sbar_{IJ}\) is an expectation value, whereas \(d_I\) is an
expectation value divided by \(c(\phi)\). The Duhamel representation gives
\begin{equation*}
    [E,\boldsymbol{\hat{\sigma}}_I]
    =
    -\int_0^\tau
    e^{-(\tau-s)\boldsymbol{\hat h}_k}
    [\boldsymbol{\hat h}_k,\boldsymbol{\hat{\sigma}}_I]
    e^{-s\boldsymbol{\hat h}_k}\,ds,
\end{equation*}
so
\begin{equation*}
    \|\boldsymbol{\hat K}_I\|_{\rm op}
    \le
    \tau e^{\tau H}C_I,
    \qquad
    e^{-\tau H}\le c(\phi)\le e^{\tau H}.
\end{equation*}
Consequently,
\begin{equation*}
    \|\vec d(\phi)\|_2
    \le
    \frac12 e^{2\tau H}C_{\mathcal P}
    \le
    \sqrt M\,H e^{2\tau H},
\end{equation*}
so an admissible choice is
\(B_d=\frac12e^{2\tau H}C_{\mathcal P}\).

Finally, with
\(q(\phi):=\bra\phi E^2\ket\phi=c(\phi)^2\),
Eq.~\eqref{eq:expectation_lipschitz} implies
\begin{align*}
    |q(\phi)-q(\varphi)|
    &\le 2e^{2\tau H}\delta,
    \\
    |c(\phi)-c(\varphi)|
    &\le e^{3\tau H}\delta,
    \\
    \left|c(\phi)^{-1}-c(\varphi)^{-1}\right|
    &\le e^{5\tau H}\delta.
\end{align*}
Using also
\(
|\bra\phi\boldsymbol{\hat K}_I\ket\phi
-\bra\varphi\boldsymbol{\hat K}_I\ket\varphi|
\le2\tau e^{\tau H}C_I\delta
\), one obtains
\begin{align*}
    |d_I(\phi)-d_I(\varphi)|
    &\le
    C_I\left(e^{2\tau H}+\frac12e^{6\tau H}\right)\delta,
    \\
    \|\vec d(\phi)-\vec d(\varphi)\|_2
    &\le
    C_{\mathcal P}
    \left(e^{2\tau H}+\frac12e^{6\tau H}\right)\delta.
\end{align*}
Thus one may take
\begin{equation*}
    L_d
    =
    C_{\mathcal P}
    \left(e^{2\tau H}+\frac12e^{6\tau H}\right)
    \le
    \sqrt M\,H\left(2e^{2\tau H}+e^{6\tau H}\right).
\end{equation*}
For a finite ordered trajectory, the constants can be made uniform in the
update index by taking the maxima of these expressions. In particular, with
\(M_*:=\max_r M^{(r)}\),
\(H_*:=\max_r\|\boldsymbol{\hat h}_{k_r}\|_{\rm op}\), and
\(C_*:=\max_r C_{\mathcal P_{r}}\), one may use
\begin{equation*}
    B_d^*=\frac12e^{2\tau H_*}C_*,
    \qquad
    L_d^*=C_*\left(e^{2\tau H_*}+\frac12e^{6\tau H_*}\right),
    \qquad
    L_S^*=2M_*.
\end{equation*}
Correspondingly, a uniform stability constant in
Eq.~\eqref{eq:L_rho_definition} is
\begin{equation*}
    L_\rho^*
    :=
    \sqrt{M_*}
    \left(
    \frac{L_d^*}{\rho}
    +
    \frac{L_S^*B_d^*}{\rho^2}
    \right).
\end{equation*}

\section{Additional Numerical Data for the TFIM}
\label{app:baseline_data}

This appendix collects detailed numerical values for the TFIM benchmarks that are omitted from the main text for conciseness. We first report the data underlying the representative evolutions discussed in Sec.~\ref{sec:baseline_results}. Table~\ref{tab:baseline_numerical_data} compares standard QITE and Sparse-QITE in the ordered and disordered phases, including the final energy, absolute energy error, ground-state fidelity, agreement with the QITE trajectory, and number of active Pauli terms. These data complement the qualitative comparison in Fig.~\ref{fig:tfim_dynamics} and confirm that the substantial reduction in generator size is achieved while maintaining high physical accuracy.

\begin{table}[H]
    \centering

    \resizebox{\columnwidth}{!}{
    \begin{tabular}{llccccccc}
        \hline\hline
        Phase &
        Method &
        \(E(\tau_f)\) &
        \(\Delta E_f\) &
        \(F_{\mathrm{GS}}(\tau_f)\) &
        \(F_{\mathrm{QITE}}(\tau_f)\) &
        \(\overline{F}_{\mathrm{QITE}}\) &
        \(N_{\mathrm{terms}}\) &
        Retained fraction \\
        \hline
        Ordered &
        QITE &
        \(-8.509061\) &
        \(2.13\times10^{-5}\) &
        \(0.999995\) &
        --- &
        --- &
        \(1\,603\,455\) &
        --- \\

        Ordered &
        Sparse-QITE &
        \(-8.506222\) &
        \(2.86\times10^{-3}\) &
        \(0.999315\) &
        \(0.999322\) &
        \(0.999719\) &
        \(48\,599\) &
        \(3.03\%\) \\
        \hline
        Disordered &
        QITE &
        \(-8.504246\) &
        \(4.84\times10^{-3}\) &
        \(0.999245\) &
        --- &
        --- &
        \(95\,584\) &
        --- \\

        Disordered &
        Sparse-QITE &
        \(-8.506449\) &
        \(2.63\times10^{-3}\) &
        \(0.999445\) &
        \(0.999081\) &
        \(0.999646\) &
        \(15\,295\) &
        \(16.0\%\) \\
        \hline\hline
    \end{tabular}
    }

    \caption{Detailed numerical results for the representative TFIM evolutions shown in Fig.~\ref{fig:tfim_dynamics}. The ordered-phase simulations use \(N=8\), \(D=6\), and \(\lambda=0.002\), while the disordered-phase simulations use \(N=8\), \(D=4\), and \(\lambda=0.0015\). Here, \(E_0\) denotes the exact ground-state energy, \(\Delta E_f=|E(\tau_f)-E_0|\), \(\overline{F}_{\mathrm{QITE}}\) denotes the fidelity with standard QITE averaged over the complete trajectory, and \(N_{\mathrm{terms}}\) is the total number of active Pauli terms accumulated over all imaginary-time steps. The retained fraction is measured relative to standard QITE. Dashes indicate quantities that are not applicable to the QITE reference rows.}
    \label{tab:baseline_numerical_data}
\end{table}

\subsection{Sparsity--Accuracy Trade-off}
\label{app:tfim_lambda_sensitivity}

To complement the sparsity--accuracy analysis presented in the main text, Table~\ref{tab:tfim_lambda_sensitivity} reports representative numerical points from the corresponding Pareto experiment for the TFIM. The selected values span strongly regularized, intermediate, and high-accuracy regimes in both the ordered and disordered phases. These data illustrate how the regularization strength \(\lambda\) controls the trade-off between generator sparsity and final-state accuracy, while also showing that Sparse-QITE can retain substantially fewer active terms than standard QITE even in regimes where the final accuracy is very close to the unregularized result.

\begin{table*}[t]
    \centering
    \begin{tabular}{lccccc}
        \hline\hline
        Phase &
        \(\lambda\) &
        \(F_{\mathrm{QITE}}(\tau_f)\) &
        \(F_{\mathrm{GS}}(\tau_f)\) &
        \(N_{\mathrm{terms}}\) &
        Retained fraction \\
        \hline
        Disordered & \(1\)                & \(0.429430\) & \(0.439001\) & \(201\)      & \(0.21\%\) \\
        Disordered & \(0.03\)             & \(0.996446\) & \(0.997469\) & \(6\,092\)   & \(6.37\%\) \\
        Disordered & \(0.0015\)           & \(0.999081\) & \(0.999445\) & \(15\,295\)  & \(16.0\%\) \\
        Disordered & \(10^{-5}\)          & \(0.999996\) & \(0.999236\) & \(86\,242\)  & \(90.2\%\) \\
        Disordered & QITE \((\lambda=0)\) & ---          & \(0.999245\) & \(95\,584\)  & --- \\
        \hline
        Ordered    & \(1\)                & \(1.35\times10^{-7}\) & \(1.36\times10^{-7}\) & \(0\)       & \(0\%\) \\
        Ordered    & \(0.01\)             & \(0.982239\) & \(0.982255\) & \(22\,047\)  & \(1.37\%\) \\
        Ordered    & \(0.002\)            & \(0.999322\) & \(0.999315\) & \(48\,599\)  & \(3.03\%\) \\
        Ordered    & \(0.001\)            & \(0.999993\) & \(0.999987\) & \(121\,081\) & \(7.55\%\) \\
        Ordered    & QITE \((\lambda=0)\) & ---          & \(0.999995\) & \(1\,603\,455\) & --- \\
        \hline\hline
    \end{tabular}
    \caption{
    Representative points along the sparsity--accuracy trade-off for the TFIM. \(N_{\mathrm{terms}}\) denotes the total number of active Pauli terms accumulated over the complete trajectory, and the retained fraction is measured relative to standard QITE. The rows span strongly regularized, intermediate, and high-accuracy regimes, including solutions that approach the ground-state accuracy of standard QITE while retaining substantially fewer terms. Dashes indicate quantities that are not applicable to the QITE reference rows.}
    \label{tab:tfim_lambda_sensitivity}
\end{table*}

\subsection{Scaling with the QITE Domain Size}
\label{app:tfim_scaling}

We report here the detailed numerical data underlying the TFIM scalability analysis presented in Sec.~\ref{sec:scalability}. For each domain size \(D\) and prescribed final-state fidelity threshold \(F_{\mathrm{QITE}}\), we list the largest tested regularization strength \(\lambda\) satisfying the threshold, together with the resulting final ground-state infidelity and the total number of active Pauli terms accumulated over the complete evolution. The corresponding standard-QITE ground-state infidelity and term count are also included for reference. The ordered-phase results are reported in Table~\ref{tab:tfim_scaling_ordered}, while the disordered-phase results are given in Table~\ref{tab:tfim_scaling_disordered}.

\begin{table*}[t]
\centering
\small
\setlength{\tabcolsep}{4pt}

\begin{tabular}{ccccccc}
\hline\hline
\(D\) &
\(F_{\mathrm{QITE}}^{\mathrm{th}}\) &
\(\lambda\) &
\(1-F_{\mathrm{GS}}\) &
\(N_{\mathrm{terms}}\) &
\(1-F_{\mathrm{GS}}^{\mathrm{QITE}}\) &
\(N_{\mathrm{terms}}^{\mathrm{QITE}}\) \\
\hline

2 & \(0.95\)    & \(2.00\times10^{-1}\) & \(6.69\times10^{-1}\) & \(6.73\times10^{2}\) & \(6.60\times10^{-1}\) & \(4.80\times10^{3}\) \\
2 & \(0.99\)    & \(5.00\times10^{-2}\) & \(6.66\times10^{-1}\) & \(9.51\times10^{2}\) & \(6.60\times10^{-1}\) & \(4.80\times10^{3}\) \\
2 & \(0.999\)   & \(5.00\times10^{-3}\) & \(6.61\times10^{-1}\) & \(3.09\times10^{3}\) & \(6.60\times10^{-1}\) & \(4.80\times10^{3}\) \\
2 & \(0.9999\)  & \(3.56\times10^{-3}\) & \(6.61\times10^{-1}\) & \(3.52\times10^{3}\) & \(6.60\times10^{-1}\) & \(4.80\times10^{3}\) \\
2 & \(0.99999\) & \(2.08\times10^{-3}\) & \(6.60\times10^{-1}\) & \(3.79\times10^{3}\) & \(6.60\times10^{-1}\) & \(4.80\times10^{3}\) \\
\hline

3 & \(0.95\)    & \(2.00\times10^{-1}\) & \(3.97\times10^{-1}\) & \(3.36\times10^{3}\) & \(3.81\times10^{-1}\) & \(2.23\times10^{4}\) \\
3 & \(0.99\)    & \(2.00\times10^{-2}\) & \(3.78\times10^{-1}\) & \(1.09\times10^{4}\) & \(3.81\times10^{-1}\) & \(2.23\times10^{4}\) \\
3 & \(0.999\)   & \(3.00\times10^{-3}\) & \(3.80\times10^{-1}\) & \(1.39\times10^{4}\) & \(3.81\times10^{-1}\) & \(2.23\times10^{4}\) \\
3 & \(0.9999\)  & \(5.48\times10^{-4}\) & \(3.81\times10^{-1}\) & \(1.75\times10^{4}\) & \(3.81\times10^{-1}\) & \(2.23\times10^{4}\) \\
3 & \(0.99999\) & \(6.69\times10^{-5}\) & \(3.81\times10^{-1}\) & \(2.06\times10^{4}\) & \(3.81\times10^{-1}\) & \(2.23\times10^{4}\) \\
\hline

4 & \(0.95\)    & \(5.00\times10^{-3}\) & \(1.54\times10^{-1}\) & \(5.31\times10^{4}\) & \(1.51\times10^{-1}\) & \(9.56\times10^{4}\) \\
4 & \(0.99\)    & \(2.00\times10^{-3}\) & \(1.55\times10^{-1}\) & \(6.49\times10^{4}\) & \(1.51\times10^{-1}\) & \(9.56\times10^{4}\) \\
4 & \(0.999\)   & \(2.00\times10^{-5}\) & \(1.51\times10^{-1}\) & \(9.13\times10^{4}\) & \(1.51\times10^{-1}\) & \(9.56\times10^{4}\) \\
4 & \(0.9999\)  & \(0\)                 & \(1.51\times10^{-1}\) & \(9.56\times10^{4}\) & \(1.51\times10^{-1}\) & \(9.56\times10^{4}\) \\
4 & \(0.99999\) & \(0\)                 & \(1.51\times10^{-1}\) & \(9.56\times10^{4}\) & \(1.51\times10^{-1}\) & \(9.56\times10^{4}\) \\
\hline

5 & \(0.95\)    & \(5.00\times10^{-2}\) & \(3.22\times10^{-2}\) & \(9.32\times10^{3}\) & \(2.95\times10^{-5}\) & \(3.95\times10^{5}\) \\
5 & \(0.99\)    & \(7.00\times10^{-3}\) & \(6.96\times10^{-3}\) & \(1.87\times10^{4}\) & \(2.95\times10^{-5}\) & \(3.95\times10^{5}\) \\
5 & \(0.999\)   & \(1.50\times10^{-3}\) & \(9.11\times10^{-4}\) & \(3.45\times10^{4}\) & \(2.95\times10^{-5}\) & \(3.95\times10^{5}\) \\
5 & \(0.9999\)  & \(8.18\times10^{-5}\) & \(2.22\times10^{-5}\) & \(2.05\times10^{5}\) & \(2.95\times10^{-5}\) & \(3.95\times10^{5}\) \\
5 & \(0.99999\) & \(8.18\times10^{-5}\) & \(2.22\times10^{-5}\) & \(2.05\times10^{5}\) & \(2.95\times10^{-5}\) & \(3.95\times10^{5}\) \\
\hline

6 & \(0.95\)    & \(5.00\times10^{-2}\) & \(4.67\times10^{-2}\) & \(1.15\times10^{4}\) & \(5.37\times10^{-6}\) & \(1.60\times10^{6}\) \\
6 & \(0.99\)    & \(5.00\times10^{-3}\) & \(7.55\times10^{-3}\) & \(2.73\times10^{4}\) & \(5.37\times10^{-6}\) & \(1.60\times10^{6}\) \\
6 & \(0.999\)   & \(2.00\times10^{-3}\) & \(6.85\times10^{-4}\) & \(4.86\times10^{4}\) & \(5.37\times10^{-6}\) & \(1.60\times10^{6}\) \\
6 & \(0.9999\)  & \(1.91\times10^{-3}\) & \(7.24\times10^{-5}\) & \(5.19\times10^{4}\) & \(5.37\times10^{-6}\) & \(1.60\times10^{6}\) \\
6 & \(0.99999\) & \(1.00\times10^{-3}\) & \(1.26\times10^{-5}\) & \(1.21\times10^{5}\) & \(5.37\times10^{-6}\) & \(1.60\times10^{6}\) \\
\hline

7 & \(0.95\)    & \(1.50\times10^{-1}\) & \(4.12\times10^{-2}\) & \(3.97\times10^{3}\) & \(5.34\times10^{-6}\) & \(6.46\times10^{6}\) \\
7 & \(0.99\)    & \(8.00\times10^{-2}\) & \(8.95\times10^{-3}\) & \(9.95\times10^{3}\) & \(5.34\times10^{-6}\) & \(6.46\times10^{6}\) \\
7 & \(0.999\)   & \(2.50\times10^{-2}\) & \(9.53\times10^{-4}\) & \(1.83\times10^{4}\) & \(5.34\times10^{-6}\) & \(6.46\times10^{6}\) \\
7 & \(0.9999\)  & \(2.50\times10^{-3}\) & \(4.03\times10^{-5}\) & \(5.75\times10^{4}\) & \(5.34\times10^{-6}\) & \(6.46\times10^{6}\) \\
7 & \(0.99999\) & \(1.73\times10^{-3}\) & \(1.54\times10^{-5}\) & \(6.96\times10^{4}\) & \(5.34\times10^{-6}\) & \(6.46\times10^{6}\) \\

\hline\hline
\end{tabular}

\caption{
Detailed numerical results for the QITE-domain-size scaling analysis in the ordered phase of the \(N=8\) TFIM. For each \(D\) and fidelity threshold \(F_{\mathrm{QITE}}^{\mathrm{th}}\), the table reports the largest tested regularization strength \(\lambda\) satisfying the threshold, the resulting final ground-state infidelity, and the total number of active Pauli terms accumulated over the trajectory. The corresponding standard-QITE ground-state infidelity and term count are included for reference. For \(D=4\), no positive value of \(\lambda\) in the tested grid satisfies the two strictest thresholds, and the \(\lambda=0\) result is therefore reported.
}
\label{tab:tfim_scaling_ordered}
\end{table*}

\begin{table*}[t]
\centering
\small
\setlength{\tabcolsep}{4pt}

\begin{tabular}{ccccccc}
\hline\hline
\(D\) &
\(F_{\mathrm{QITE}}^{\mathrm{th}}\) &
\(\lambda\) &
\(1-F_{\mathrm{GS}}\) &
\(N_{\mathrm{terms}}\) &
\(1-F_{\mathrm{GS}}^{\mathrm{QITE}}\) &
\(N_{\mathrm{terms}}^{\mathrm{QITE}}\) \\
\hline

2 & \(0.95\)    & \(2.00\times10^{-1}\) & \(6.27\times10^{-2}\) & \(5.90\times10^{2}\) & \(8.40\times10^{-2}\) & \(4.80\times10^{3}\) \\
2 & \(0.99\)    & \(5.00\times10^{-2}\) & \(7.10\times10^{-2}\) & \(9.22\times10^{2}\) & \(8.40\times10^{-2}\) & \(4.80\times10^{3}\) \\
2 & \(0.999\)   & \(1.00\times10^{-2}\) & \(8.15\times10^{-2}\) & \(1.32\times10^{3}\) & \(8.40\times10^{-2}\) & \(4.80\times10^{3}\) \\
2 & \(0.9999\)  & \(8.18\times10^{-4}\) & \(8.38\times10^{-2}\) & \(2.86\times10^{3}\) & \(8.40\times10^{-2}\) & \(4.80\times10^{3}\) \\
2 & \(0.99999\) & \(3.67\times10^{-4}\) & \(8.39\times10^{-2}\) & \(3.56\times10^{3}\) & \(8.40\times10^{-2}\) & \(4.80\times10^{3}\) \\
\hline

3 & \(0.95\)    & \(2.50\times10^{-1}\) & \(5.61\times10^{-2}\) & \(3.13\times10^{3}\) & \(2.64\times10^{-2}\) & \(2.23\times10^{4}\) \\
3 & \(0.99\)    & \(5.00\times10^{-2}\) & \(2.16\times10^{-2}\) & \(3.25\times10^{3}\) & \(2.64\times10^{-2}\) & \(2.23\times10^{4}\) \\
3 & \(0.999\)   & \(3.00\times10^{-2}\) & \(2.38\times10^{-2}\) & \(3.55\times10^{3}\) & \(2.64\times10^{-2}\) & \(2.23\times10^{4}\) \\
3 & \(0.9999\)  & \(6.69\times10^{-3}\) & \(2.65\times10^{-2}\) & \(7.14\times10^{3}\) & \(2.64\times10^{-2}\) & \(2.23\times10^{4}\) \\
3 & \(0.99999\) & \(4.48\times10^{-4}\) & \(2.63\times10^{-2}\) & \(1.05\times10^{4}\) & \(2.64\times10^{-2}\) & \(2.23\times10^{4}\) \\
\hline

4 & \(0.95\)    & \(2.50\times10^{-1}\) & \(4.72\times10^{-2}\) & \(4.20\times10^{3}\) & \(7.55\times10^{-4}\) & \(9.56\times10^{4}\) \\
4 & \(0.99\)    & \(5.00\times10^{-2}\) & \(5.02\times10^{-3}\) & \(6.34\times10^{3}\) & \(7.55\times10^{-4}\) & \(9.56\times10^{4}\) \\
4 & \(0.999\)   & \(1.50\times10^{-3}\) & \(5.55\times10^{-4}\) & \(1.53\times10^{4}\) & \(7.55\times10^{-4}\) & \(9.56\times10^{4}\) \\
4 & \(0.9999\)  & \(3.67\times10^{-5}\) & \(8.32\times10^{-4}\) & \(7.31\times10^{4}\) & \(7.55\times10^{-4}\) & \(9.56\times10^{4}\) \\
4 & \(0.99999\) & \(1.73\times10^{-5}\) & \(7.66\times10^{-4}\) & \(8.25\times10^{4}\) & \(7.55\times10^{-4}\) & \(9.56\times10^{4}\) \\
\hline

5 & \(0.95\)    & \(2.50\times10^{-1}\) & \(4.51\times10^{-2}\) & \(3.51\times10^{3}\) & \(1.46\times10^{-4}\) & \(3.95\times10^{5}\) \\
5 & \(0.99\)    & \(5.00\times10^{-2}\) & \(3.31\times10^{-3}\) & \(8.06\times10^{3}\) & \(1.46\times10^{-4}\) & \(3.95\times10^{5}\) \\
5 & \(0.999\)   & \(1.50\times10^{-2}\) & \(8.66\times10^{-4}\) & \(9.78\times10^{3}\) & \(1.46\times10^{-4}\) & \(3.95\times10^{5}\) \\
5 & \(0.9999\)  & \(8.18\times10^{-4}\) & \(1.80\times10^{-4}\) & \(2.83\times10^{4}\) & \(1.46\times10^{-4}\) & \(3.95\times10^{5}\) \\
5 & \(0.99999\) & \(3.67\times10^{-4}\) & \(1.51\times10^{-4}\) & \(4.89\times10^{4}\) & \(1.46\times10^{-4}\) & \(3.95\times10^{5}\) \\
\hline

6 & \(0.95\)    & \(2.50\times10^{-1}\) & \(4.29\times10^{-2}\) & \(3.34\times10^{3}\) & \(7.10\times10^{-5}\) & \(1.60\times10^{6}\) \\
6 & \(0.99\)    & \(1.00\times10^{-1}\) & \(9.13\times10^{-3}\) & \(1.02\times10^{4}\) & \(7.10\times10^{-5}\) & \(1.60\times10^{6}\) \\
6 & \(0.999\)   & \(2.00\times10^{-2}\) & \(7.52\times10^{-4}\) & \(1.38\times10^{4}\) & \(7.10\times10^{-5}\) & \(1.60\times10^{6}\) \\
6 & \(0.9999\)  & \(1.44\times10^{-3}\) & \(1.59\times10^{-4}\) & \(3.44\times10^{4}\) & \(7.10\times10^{-5}\) & \(1.60\times10^{6}\) \\
6 & \(0.99999\) & \(6.69\times10^{-4}\) & \(7.96\times10^{-5}\) & \(7.78\times10^{4}\) & \(7.10\times10^{-5}\) & \(1.60\times10^{6}\) \\
\hline

7 & \(0.95\)    & \(2.50\times10^{-1}\) & \(4.24\times10^{-2}\) & \(3.62\times10^{3}\) & \(7.05\times10^{-5}\) & \(6.46\times10^{6}\) \\
7 & \(0.99\)    & \(1.00\times10^{-1}\) & \(8.81\times10^{-3}\) & \(1.35\times10^{4}\) & \(7.05\times10^{-5}\) & \(6.46\times10^{6}\) \\
7 & \(0.999\)   & \(2.00\times10^{-2}\) & \(7.37\times10^{-4}\) & \(1.76\times10^{4}\) & \(7.05\times10^{-5}\) & \(6.46\times10^{6}\) \\
7 & \(0.9999\)  & \(6.69\times10^{-3}\) & \(1.57\times10^{-4}\) & \(3.15\times10^{4}\) & \(7.05\times10^{-5}\) & \(6.46\times10^{6}\) \\
7 & \(0.99999\) & \(6.69\times10^{-4}\) & \(7.67\times10^{-5}\) & \(2.37\times10^{5}\) & \(7.05\times10^{-5}\) & \(6.46\times10^{6}\) \\

\hline\hline
\end{tabular}

\caption{
Detailed numerical results for the QITE-domain-size scaling analysis in the disordered phase of the \(N=8\) TFIM. For each \(D\) and fidelity threshold \(F_{\mathrm{QITE}}^{\mathrm{th}}\), the table reports the largest tested regularization strength \(\lambda\) satisfying the threshold, the resulting final ground-state infidelity, and the total number of active Pauli terms accumulated over the trajectory. The corresponding standard-QITE ground-state infidelity and term count are included for reference.
}
\label{tab:tfim_scaling_disordered}
\end{table*}

\section{Additional Results for the Heisenberg Model}
\label{app:heisenberg_results}

In this appendix, we provide additional numerical results for the antiferromagnetic Heisenberg model that complement the representative evolution presented in Fig.~\ref{fig:tfim_dynamics}. For conciseness, we report only the detailed numerical values associated with this representative evolution. The corresponding sparsity--accuracy and domain-size scaling experiments are not shown, as they display trends qualitatively similar to those observed for the TFIM.

The detailed numerical values corresponding to the representative Heisenberg evolution shown in Fig.~\ref{fig:tfim_dynamics} are reported in Table~\ref{tab:heisenberg_numerical_data}. We compare standard QITE and Sparse-QITE for \(N=8\), \(D=6\), and \(\lambda=0.03\). Over the complete evolution, Sparse-QITE reduces the total number of active Pauli terms from \(788\,928\) to \(46\,588\), corresponding to \(5.91\%\) of the standard-QITE total. At the same time, both the final and mean fidelities with standard QITE remain above \(0.999\), while the final ground-state fidelity remains close to that obtained with standard QITE.

\begin{table}[H]
    \centering

    \resizebox{\columnwidth}{!}{
    \begin{tabular}{lcccccccc}
        \hline\hline
        Method &
        \(E(\tau_f)\) &
        \(\Delta E_f\) &
        \(F_{\mathrm{GS}}(\tau_f)\) &
        \(F_{\mathrm{QITE}}(\tau_f)\) &
        \(\overline{F}_{\mathrm{QITE}}\) &
        \(N_{\mathrm{terms}}\) &
        Retained fraction \\
        \hline
        QITE &
        \(-14.580891\) &
        \(2.35\times10^{-2}\) &
        \(0.997156\) &
        --- &
        --- &
        \(788\,928\) &
        --- \\

        Sparse-QITE &
        \(-14.573858\) &
        \(3.05\times10^{-2}\) &
        \(0.996287\) &
        \(0.999449\) &
        \(0.999269\) &
        \(46\,588\) &
        \(5.91\%\) \\
        \hline\hline
    \end{tabular}
    }

    \caption{
    Detailed numerical results for the antiferromagnetic Heisenberg evolution shown in Fig.~\ref{fig:tfim_dynamics}. The simulations use \(N=8\), \(D=6\), and \(\lambda=0.03\). Here, \(E_0\) denotes the exact ground-state energy, \(\Delta E_f=|E(\tau_f)-E_0|\), \(\overline{F}_{\mathrm{QITE}}\) denotes the fidelity with standard QITE averaged over the complete trajectory, and \(N_{\mathrm{terms}}\) is the total number of active Pauli terms accumulated over all imaginary-time steps. The retained fraction is measured relative to standard QITE. Dashes indicate quantities that are not applicable to the QITE reference row.
    }
    \label{tab:heisenberg_numerical_data}
\end{table}

\section{Circuit Compilation}
\label{app:compilation_scalability}

To synthesize the QITE generators, we use the symmetric second-order
Suzuki product formula. For a generator
$A=\sum_{j=1}^{M} P_j$, one repetition approximates its exponential as
\[
e^{-i\Delta\tau A}
\approx
\left(\prod_{j=1}^{M} e^{-i\Delta\tau \boldsymbol{\hat{P}}_j/2}\right)
\left(\prod_{j=M}^{1} e^{-i\Delta\tau \boldsymbol{\hat{P}}_j/2}\right).
\]
The forward--backward symmetry cancels the leading error of the
first-order Lie--Trotter formula, producing a global error that scales
approximately as $\mathcal{O}(\Delta\tau^3/r^2)$ when the formula is
applied using $r$ repetitions.

The repetition count is chosen independently for each generator using
the heuristic rule
\[
r_{n,k}
=
\max\left\{
1,
\left\lceil
\frac{\Delta\tau\max_j |a_{n,k,j}|}{\theta_{\max}}
\right\rceil
\right\},
\]
where $a_{n,k,j}$ is the coefficient of the $j$th Pauli term in the
$k$th local generator at QITE step $n$. This rule keeps the largest
coefficient angle in each internal product-formula step below the
prescribed value $\theta_{\max}$. Consequently, generators with
moderate coefficients retain a single repetition, whereas generators
containing unusually large coefficients are divided into smaller
steps. This criterion is a practical stability heuristic rather than
a rigorous error bound, since it does not explicitly include
commutators between Pauli terms.

We now focus on the resulting circuit resources and on whether the sparsity observed at the generator level is preserved after compilation. As expected, the compiled circuits follow the same qualitative trend as the number of active generator terms: reducing the Sparse-QITE support translates into substantial reductions in both gate count and circuit depth, with the savings becoming particularly pronounced at larger domain sizes.

For conciseness, Table~\ref{tab:tfim_compilation_scalability} reports two representative fidelity thresholds, \(F_{\mathrm{QITE}}\geq0.999\) and \(F_{\mathrm{QITE}}\geq0.99999\), and focuses on the domain sizes for which QITE already provides an accurate approximation to the ground state, namely \(D\geq4\) in the disordered phase and \(D\geq5\) in the ordered phase.

\begin{table*}[t]
\centering
\small
\setlength{\tabcolsep}{4pt}

\begin{tabular}{lcccccc}
\hline\hline
Phase &
\(D\) &
\(F_{\mathrm{QITE}}^{\mathrm{th}}\) &
U3 gates
(SQ/QITE) &
CX gates
(SQ/QITE) &
Depth
(SQ/QITE) &
\(F_{\mathrm{GS}}^{\mathrm{comp}}\) \\
\hline

Disordered
& 4 & \(0.999\)
& \(6.62\times10^{4}/4.03\times10^{5}\)
& \(7.40\times10^{4}/4.57\times10^{5}\)
& \(1.17\times10^{5}/7.52\times10^{5}\)
& \(0.999446\) \\

Disordered
& 5 & \(0.999\)
& \(3.79\times10^{4}/1.70\times10^{6}\)
& \(4.33\times10^{4}/1.96\times10^{6}\)
& \(6.58\times10^{4}/3.20\times10^{6}\)
& \(0.999133\) \\

Disordered
& 6 & \(0.999\)
& \(5.71\times10^{4}/6.93\times10^{6}\)
& \(7.52\times10^{4}/8.03\times10^{6}\)
& \(1.11\times10^{5}/1.31\times10^{7}\)
& \(0.999248\) \\

Disordered
& 7 & \(0.999\)
& \(7.19\times10^{4}/2.80\times10^{7}\)
& \(1.07\times10^{5}/3.24\times10^{7}\)
& \(1.53\times10^{5}/5.28\times10^{7}\)
& \(0.999262\) \\
\hline

Disordered
& 4 & \(0.99999\)
& \(3.51\times10^{5}/4.03\times10^{5}\)
& \(3.99\times10^{5}/4.57\times10^{5}\)
& \(6.51\times10^{5}/7.52\times10^{5}\)
& \(0.999234\) \\

Disordered
& 5 & \(0.99999\)
& \(1.99\times10^{5}/1.70\times10^{6}\)
& \(2.70\times10^{5}/1.96\times10^{6}\)
& \(4.05\times10^{5}/3.20\times10^{6}\)
& \(0.999848\) \\

Disordered
& 6 & \(0.99999\)
& \(3.56\times10^{5}/6.93\times10^{6}\)
& \(5.04\times10^{5}/8.03\times10^{6}\)
& \(7.33\times10^{5}/1.31\times10^{7}\)
& \(0.999870\) \\

Disordered
& 7 & \(0.99999\)
& \(1.11\times10^{6}/2.80\times10^{7}\)
& \(1.76\times10^{6}/3.24\times10^{7}\)
& \(2.44\times10^{6}/5.28\times10^{7}\)
& \(0.999877\) \\
\hline\hline

Ordered
& 5 & \(0.999\)
& \(1.55\times10^{5}/1.70\times10^{6}\)
& \(2.11\times10^{5}/1.96\times10^{6}\)
& \(3.10\times10^{5}/3.20\times10^{6}\)
& \(0.999087\) \\

Ordered
& 6 & \(0.999\)
& \(2.24\times10^{5}/6.93\times10^{6}\)
& \(3.30\times10^{5}/8.03\times10^{6}\)
& \(4.67\times10^{5}/1.31\times10^{7}\)
& \(0.999316\) \\

Ordered
& 7 & \(0.999\)
& \(8.53\times10^{4}/2.80\times10^{7}\)
& \(1.25\times10^{5}/3.24\times10^{7}\)
& \(1.76\times10^{5}/5.28\times10^{7}\)
& \(0.999043\) \\
\hline

Ordered
& 5 & \(0.99999\)
& \(9.35\times10^{5}/1.70\times10^{6}\)
& \(1.18\times10^{6}/1.96\times10^{6}\)
& \(1.81\times10^{6}/3.20\times10^{6}\)
& \(0.999674\) \\

Ordered
& 6 & \(0.99999\)
& \(5.98\times10^{5}/6.93\times10^{6}\)
& \(8.78\times10^{5}/8.03\times10^{6}\)
& \(1.25\times10^{6}/1.31\times10^{7}\)
& \(0.999477\) \\

Ordered
& 7 & \(0.99999\)
& \(3.38\times10^{5}/2.80\times10^{7}\)
& \(5.28\times10^{5}/3.24\times10^{7}\)
& \(7.32\times10^{5}/5.28\times10^{7}\)
& \(0.999801\) \\

\hline\hline
\end{tabular}

\caption{Compiled circuit resources for representative Sparse-QITE scalability results in the \(N=8\) TFIM. We report the thresholds \(F_{\mathrm{QITE}}^{\mathrm{th}}=0.999\) and \(0.99999\), restricting the comparison to \(D\geq4\) in the disordered phase and \(D\geq5\) in the ordered phase. U3 gate counts, CX gate counts, and circuit depths are shown as Sparse-QITE/QITE. \(F_{\mathrm{GS}}^{\mathrm{comp}}\) denotes the ground-state fidelity obtained after compilation of the Sparse-QITE evolution.
}
\label{tab:tfim_compilation_scalability}
\end{table*}


For the intermediate threshold \(F_{\mathrm{QITE}}\geq0.999\), the reduction in circuit resources is particularly strong. In the disordered phase, both the U3 and CX gate counts decrease from roughly \(16\%\) of their standard-QITE values at \(D=4\) to below \(1\%\) at \(D=6\) and \(7\). A similar trend is observed in the ordered phase, where the reductions become increasingly pronounced with \(D\), reaching well below \(1\%\) of the standard-QITE gate counts at \(D=7\). The circuit depth is reduced by a comparable amount. Thus, the mild growth of the Sparse-QITE active support observed in the scalability analysis translates directly into very large reductions in both one- and two-qubit circuit resources.

Importantly, substantial savings remain even for the much stricter threshold \(F_{\mathrm{QITE}}\geq0.99999\). In the disordered phase, the U3 and CX gate counts at the largest domain sizes remain only a few percent of their standard-QITE values. In the ordered phase, the reduction is less pronounced at \(D=5\), but becomes increasingly strong as \(D\) grows, reaching approximately \(1\%\)--\(2\%\) of the standard-QITE gate counts at \(D=7\). These results show that the compression observed at the generator level survives the circuit decomposition and transpilation, yielding substantial reductions in gate counts and circuit depth even in the high-fidelity regime.

The corresponding scalability analysis and circuit-compilation results for the antiferromagnetic Heisenberg model are presented in App.~\ref{app:heisenberg_results}, where we apply the same domain-size and fidelity-threshold analysis used for the TFIM.

\end{document}